\documentclass[a4paper,twoside]{aa}

\usepackage{aalongtable}
\usepackage{amsbsy}
\usepackage{amstext}
\usepackage[final]{graphicx}
\usepackage{natbib}
\usepackage{txfonts}

\begin{document}

\title{HE~0515--4414 -- an unusual sub-damped Ly~$\alpha$ system revisited%
  \thanks{Based on observations made with ESO Telescopes at the La Silla
    or Paranal Observatories under programme ID~066.A-0212.}
  \fnmsep
  \thanks{Based on observations made with the NASA/ESA Hubble Space
    Telescope, obtained from the data archive at the Space Telescope
    Institute. STScI is operated by the association of Universities for
    Research in Astronomy, Inc. under the NASA contract NAS~5-26555.}}
\subtitle{}
\author{R. Quast\inst1 \and
  D. Reimers\inst2 \and
  R. Baade\inst2}
\titlerunning{HE~0515--4414 -- an unusual sub-damped Ly~$\alpha$ system
  revisited}
\authorrunning{R.~Quast et al.}
\institute{%
  Brockmann Consult, 
  GKSS-Forschungszentrum,
  Max-Planck-Stra\ss e 2,
  D-21502 Geesthacht,
  Germany
  \and
  Hamburger Sternwarte,
  Universit\"at Hamburg,
  Gojenbergsweg 112,
  D-21029 Hamburg,
  Germany}
\date{accepted}

\abstract
{}
{We examine the ionization, abundances, and differential dust depletion of
metals, the kinematic structure, and the physical conditions in the
molecular hydrogen-bearing sub-damped Ly~$\alpha$ system toward
HE~0515-4414.}
{We used the STIS and VLT UVES spectrographs to obtain high-resolution 
recordings of the damped Ly~$\alpha$ profile and numerous associated 
metal lines. Observed element abundances are corrected with respect to 
dust depletion effects.}
{The sub-damped Ly~$\alpha$ absorber at redshift $z=1.15$ is unusual in 
several aspects. The velocity interval of associated metal lines extends 
for $700$~km\,s$^{-1}$. 
In addition, saturated \ion{H}{i} absorption is detected in the blue damping
wing of the $N_\mathrm{\ion{H}{i}}=8\times10^{19}$~cm$^{-2}$ main component.
The column density ratios of associated \ion{Al}{ii}, \ion{Al}{iii}, and
\ion{Fe}{ii} lines indicate that the absorbing material is ionized. 19 of in
total 31 detected metal line components are formed within peripheral 
\ion{H}{ii} regions, while only 12 components are associated with the
predominantly neutral main absorber. 
The bimodal velocity distribution of metal line components suggests two 
interacting absorbers. For the main absorber the observed abundance
ratios of refractory elements to Zn range from 
Galactic warm disk $[\mathrm{Si}/\mathrm{Zn}]_\mathrm{g}=-0.40\pm0.06$, 
$[\mathrm{Fe}/\mathrm{Zn}]_\mathrm{g}=-1.10\pm0.05$ to halo-like and 
essentially undepleted patterns. The dust-corrected metal abundances 
indicate a nucleosynthetic odd-even effect and might imply an anomalous 
depletion of Si relative to Fe for two components, but otherwise do 
correspond to solar ratios. The intrinsic average metallicity is almost 
solar $[\mathrm{Fe}/\mathrm{H}]_\mathrm{m}=-0.08\pm0.19$, whereas the 
uncorrected average is 
$[\mathrm{Zn}/\mathrm{H}]_\mathrm{g}=-0.38\pm0.04$. The ion abundances in 
the periphery conform with solar element composition.}
{The  
detection of \ion{H}{ii} as well as the large variation in dust
depletion for this sight line raises the question whether in future studies
of damped Ly~$\alpha$ systems ionization and depletion effects have to be
considered in further detail. Ionization effects, for instance, may pretend
an enrichment of $\alpha$ elements. An empirical recipe for detecting 
\ion{H}{ii} regions is provided.}

\keywords{cosmology: observations -
  galaxies: abundances --
  galaxies: interactions --
  intergalactic medium --
  quasars: absorption lines --
  quasars: individual: HE~0515--4414}

\maketitle

\section{Introduction} 

The study of QSO absorption lines provides vital information on the 
nucleosynthetic history of the universe by complementing the 
compositional analysis of stars and interstellar space in local galaxies 
with element abundances at higher redshift. In particular, interests are 
focused on extragalactic structures termed damped Ly $\alpha$ (DLA) 
systems, essentially comprised of neutral hydrogen with column densities 
$N_\ion{H}{i}\ge2\times10^{20}$ atoms cm$^{-2}$ \citep[for a review 
see][]{WolfeGP_2005}. Absorbers in the sub-DLA range with column 
densities $N_\ion{H}{i}\ge10^{19}$ atoms cm$^{-2}$ might be mainly 
neutral when the ionizing background is reduced 
\citep{PerouxDKMD_2002,PerouxDDKM_2003}. The aim of these examinations is 
to establish accurate element abundances for the aggregations of neutral 
gas that are examples of interstellar environments in the high-redshift 
universe. Since the measurement of metal column densities is 
straightforward, the only problem is their correct interpretation.

The true nature of DLA systems is unknown and the underlying population, 
being constituted of hierarchical structures with different 
morphologies, chemical enrichment histories, and physical environments, 
is multifarious. The diversity is attested by the disparate values 
obtained for metal abundances at any given redshift.

The metallicity of DLA systems is not correlated with their column 
density, 
however, there is an upper bound for distribution of column densities versus
metallicity \citep{BoisseLBD_1998}.
Though there are high column density DLA absorbers with high 
metallicity in the foreground of the star forming hosts of gamma-ray 
bursts \citep{Watson_2005}, similar absorbers are not detected 
toward QSOs. The cosmic mean metallicity of DLA absorbers increases with 
cosmic time \citep{ProchaskaGWCD_2003,KulkarniFL_2005,RaoPHW_2005}, but 
is an order of magnitude lower than predicted by cosmic star formation 
history \citep[see the discussion of the missing metals problem 
by][]{WolfeGP_2005}. The solution to this problem is a matter of debate. 
Conclusive evidence of enriched material ejected from DLA absorbers into 
the intergalactic medium or of active star formation restricted to 
compact regions is missing. The latter possibility is closely linked to 
the physical properties of the interstellar medium and its molecular 
content \citep{WolfePG_2003,WolfeGP_2003}. Molecular gas is uncommon in 
DLA absorbers. If found, the fraction of molecular hydrogen, usually 
between $10^{-6}$ and $10^{-2}$, is not correlated with the column 
density of atomic hydrogen \citep{LedouxPS_2003}. 
However, \citet{PetitjeanLNS_2006} have demonstrated that the presence
of molecular hydrogen at high redshift is strongly correlated with the
metallicity.
 
Since the spectroscopic analysis is restricted to the gaseous phase of 
the absorbing medium, observed element abundances are potentially 
distorted by dust removing atoms in varying amounts, depending on their 
affinity to the solid state. In particular high-metallicity and 
molecule-bearing absorbers are affected by dust 
\citep{PetitjeanSL_2002,LedouxPS_2003}. Depletions are largely lower than 
in the Galactic halo, but increase with metallicity 
\citep{VladiloG_2004}. In practice, the observed element abundances are 
corrected ad hoc, using Galactic interstellar depletion patterns as 
reference \citep{VladiloG_2002a,VladiloG_2002b}. Another aspect of dust 
is the possibility that DLA absorbers may elude detection because the 
background QSOs are obscured \citep{FallS_PeiY_1993}. The selection 
effects are complicated since obscurement is counteracted, but not 
compensated, by amplification due to gravitational lensing 
\citep{SmetteCS_1997}. The effect of dust is subject of several studies 
\citep{MurphyM_LiskeJ_2004,QuastRS_2004,AkermanEPS_2005,SmetteWL_2005,VladiloG_PerouxC_2005,WildHP_2005}. 
A further difficulty are ionization effects. Examples of DLA-associated 
metal line components formed within mainly ionized material are given by 
\citet{ProchaskaHO_2002} and \citet{DessaugesPD_2006}.

The column density distribution and kinematic structure of absorbers 
provide important constraints on hierarchical structuring 
\citep[e.g.][]{CenOPW_2003,NagamineSH_2004} and immediate insight into 
the processes of galaxy formation \citep{WolfeA_ProchaskaJ_2000a}. The 
extended multicomponent velocity structure and characteristic asymmetry 
of DLA-associated metal lines is consistent with galaxy formation models 
in hierarchic cold dark matter cosmologies, and reproducible by the 
hydrodynamical simulation of rotation, random motion, infall, and merging 
of irregular protogalactic clumps hosted by collapsed dark matter halos 
\citep{HaehneltSR_1998}. The velocity structure of sub-DLA absorbers 
compares to that of the higher column density systems 
\citep{PerouxDDKM_2003}, which is unexpected since semianalytic galaxy 
formation models \citep{MallerPSP_2001,MallerPSP_2003} predict markedly 
different kinematic properties. The absorption velocity intervals of both 
sub-DLA and DLA absorbers typically extend for $100$~km\,s$^{-1}$. More 
extended systems tend to higher metallicities and lower hydrogen column 
densities \citep{WolfeA_ProchaskaJ_1998}. In particular the latter property is 
unexpected and difficult to interpret in terms of rotating disks models. 
The most extended systems, however, are probably due to interacting or 
merging galaxies \citep{PetitjeanSL_2002,RichterLPB_2005}. 
Strong observational 
evidence for a correlation between DLA metallicity and absorption profile
velocity spread, which probably is the consequence of a mass-metalicity
relation, has recently been provided by \citet{LedouxPFMS_2006}.

In this study we revisit the $z=1.15$ sub-DLA system toward HE~0515--4414 
\citep{ReimersHRW_1998,VargaRTBB_2000}. The main components of associated 
metal lines exhibit excited neutral carbon and molecular hydrogen 
\citep{QuastBR_2002,ReimersBQL_2003}. Most outstanding, the absorption 
velocity interval extends for $700$~km\,s$^{-1}$. Based on refined 
spectroscopy, we examine the ionization, abundances, and differential dust 
depletion of metals as well as the kinematic structure, and physical
conditions of this unusual absorption line system.

\section{Observations} 

\begin{figure*} \centering
\includegraphics{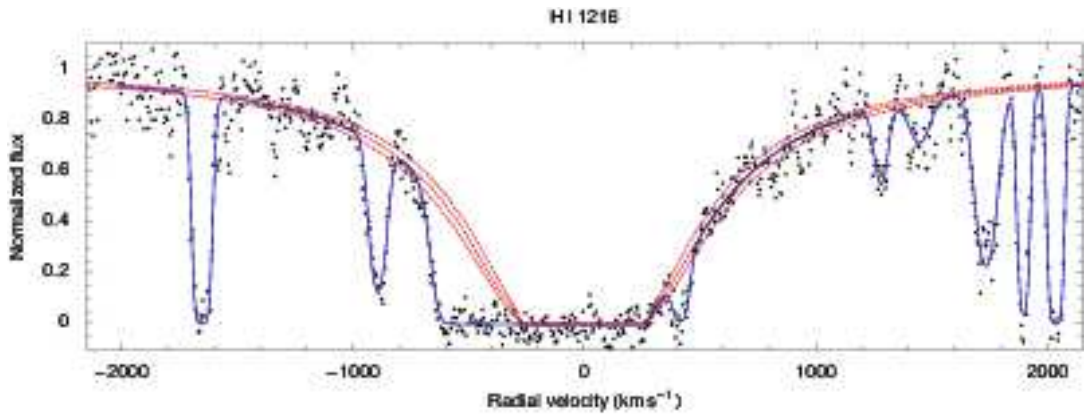}
\caption[]{STIS Echelle order showing the sub-damped \ion{H}{i} profile
with redshift $z=1.15$. The solid curve indicates the optimized profile 
decomposition of the spectrum, while the dashed curves mark the damped
profile. The blue damping wing is blended with further \ion{H}{i} absorption 
which is associated with numerous metal lines (Fig.~\ref{fg:metals}). The
origin of the radial velocity axis corresponds to the redshift $z=1.15080$}
\label{fg:dla}
\end{figure*}

\begin{figure*} \centering
\includegraphics{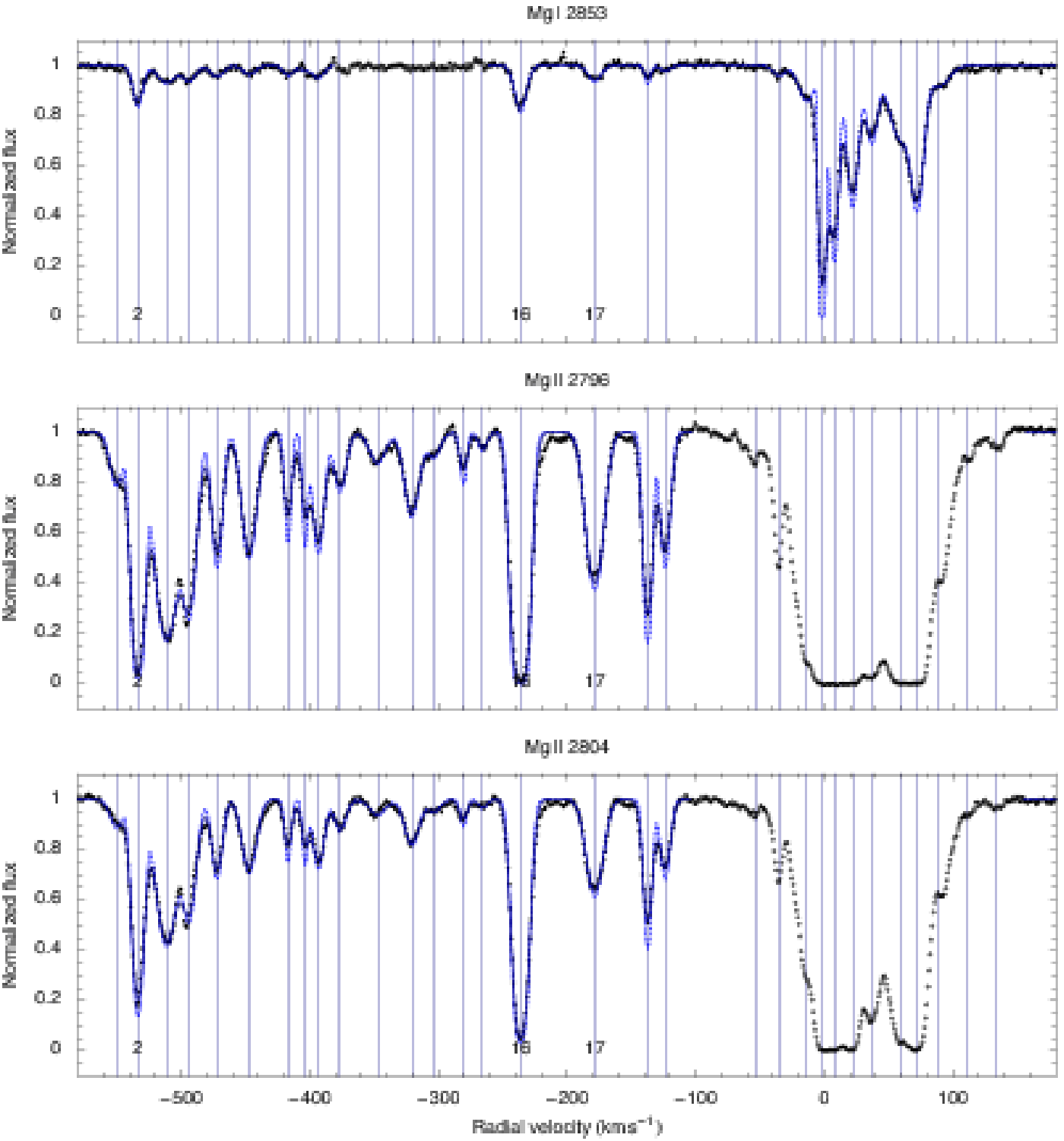}
\caption[]{Metal lines associated with the sub-damped profile shown in 
Fig.~\ref{fg:dla}. The dashed and solid curves indicate the optimized 
profile decomposition and its instrumental convolution. Individual 
components are marked by vertical lines. Data represented by empty 
circles are ignored in the profile decomposition. Components 3, 16, and 
17 possibly are unresolved blends \citep{QuastRL_2004}. This figure is
continued in the Online Material}
\label{fg:metals}
\end{figure*}

Ranging from the NUV to the end of the visual, the observations cover the
sub-damped profile at 2615~\AA\ (Fig.~\ref{fg:dla}) and numerous 
associated metal lines (Fig.~\ref{fg:metals}).

\begin{table} \centering
\caption[]{Details of spectra obtained with UVES}
\begin{tabular}{@{}llllll@{}}
\hline
\hline
Date\rule[-5pt]{0pt}{15pt}
& Obs.   & Exp. (s) & Mode & Arm  & Wav. (\AA) \\
\hline
2000-10-07 & 101818 & 4500 & DI2  & blue & 3460\rule{0pt}{10pt} \\
           &        & 4499 &      & red  & 8600 \\
2000-11-16 & 101822 & 4500 & DI1  & blue & 3460 \\
           &        & 4499 &      & red  & 5800 \\
2000-11-17 & 101821 & 4500 & DI1  & blue & 3460 \\
           &        & 4499 &      & red  & 5800 \\
2000-11-18 & 101820 & 4500 & DI1  & blue & 3460 \\
           &        & 4499 &      & red  & 5800 \\
2000-12-15 & 101812 & 3600 & DI2  & blue & 4370 \\
           &        & 3599 &      & red  & 8600 \\
           & 101813 & 3600 & DI2  & blue & 4370 \\
           &        & 3600 &      & red  & 8600 \\
           & 101814 & 3600 & DI2  & blue & 4370 \\
           &        & 3599 &      & red  & 8600 \\
2000-12-16 & 101811 & 3600 & DI2  & blue & 4370 \\
           &        & 3600 &      & red  & 8600 \\
2000-12-21 & 101810 & 3600 & DI2  & blue & 4370 \\
           &        & 3599 &      & red  & 8600 \\
           & 101815 & 3600 & DI2  & blue & 4370 \\
           &        & 3599 &      & red  & 8600 \\
2000-12-23 & 101817 & 4500 & DI2  & blue & 3460 \\
           &        & 4500 &      & red  & 8600 \\
2000-12-24 & 101819 & 4500 & DI1  & blue & 3460 \\
           &        & 4500 &      & red  & 5800 \\
2001-01-02 & 101816 & 4500 & DI2  & blue & 3460 \\
           &        & 4500 &      & red  & 8600 \\
\hline
\end{tabular}
\label{tb:obs}
\end{table}

\subsection{UV-visual spectroscopy}
HE~0515--4414 was observed during ten nights between October~7, 2000 and 
January~3, 2001, using the UV-Visual Echelle Spectrograph (UVES) installed 
at the second VLT Unit Telescope (Kueyen). Thirteen exposures were made in 
the dichroic mode using standard settings for the central wavelengths of 
3460/4370~\AA\ in the blue, and 5800/8600~\AA\ in the red 
(Table~\ref{tb:obs}). The CCDs were read out in fast mode without binning. 
Individual exposure times were 3600 and 4500~s, under photometric to clear 
sky and seeing conditions ranging from 0.47 to 0.70 arcsec. The slit width 
was 0.8 arcsec providing a spectral resolution of about 55\,000 in the 
blue and slightly less in the red. The raw data frames were reduced at the 
ESO Quality Control Garching using the UVES pipeline Data Reduction 
Software. Finally, the individual vacuum-barycentric corrected spectra 
were combined resulting in an effective signal-to-noise ratio per pixel of 
90-140.

\subsection{NUV spectroscopy}
The UV-visual recordings were supplemented by spectra obtained with the 
Space Telescope Imaging Spectrograph (STIS) during three orbits between 
January~31 and February~1, 2000, ranging from 2300 to 3100~\AA. The total 
exposure time was 31\,500~s, while the instrument was operating in the 
medium resolution NUV mode (E230M) with the entrance aperture of 
$0\farcs2\times0\farcs2$ providing a spectral resolution of about 
30\,000. The raw spectra were reduced at the Space Telescope Science 
Institute using the STIS pipeline software completed by an additional 
interorder background correction. The combined spectra show an effective 
signal-to-noise ratio per pixel of 5-10.

\section{Line profile analysis} 

There are several basically different techniques for the analysis of QSO 
absorption lines: the classical line profile decomposition, the apparent 
optical depth method \citep{SavageB_SembachK_1991}, and Monte Carlo 
inversion \citep{LevshakovAK_2000}. While the classical profile 
decomposition postulates discrete homogenous absorbers with Gaussian 
(i.e. thermal or microturbulent) particle velocity distributions, the 
apparent optical depth technique allows the very direct interpretation of 
observed spectra without the need to consider the velocity structure of 
spectral lines as long as the absorption is optically thin or 
sufficiently resolved. Otherwise, the apparent optical depth is not 
representative and additional operations are required to recover the 
correct column density \citep{JenkinsE_1996}. The corrective procedure, 
however, is only approved for synthetic spectra with Gaussian velocity 
distributions underlying the individual components. Monte Carlo inversion 
considers random velocity and density fields along the sight line, but 
requires photoionization calculations to connect the random fields to the 
observed absorption that are too time-consuming for DLA systems.

Since we consider many blended or optically thick lines, we prefer the 
classical decomposition technique for the analysis and use the apparent 
optical depth method to supplement the diagnostics. Throughout the 
analysis we use the atomic data compiled by \citet{MortonD_2003}.
 
\subsection{Line profile decomposition}

The general problem of line profile decomposition in QSO spectra and its 
solution by means of evolutionary forward modelling is described in detail 
by \citet{QuastBR_2005}. For the specific purpose of measuring accurate 
metal column densities we introduce several additional constraints:
\begin{enumerate}
  \item Each metal line component is modelled by a superposition of 
  Doppler profiles positioned at the same radial velocity. This procedure 
  helps to recover the velocity structure of the instrumentally blurred 
  line ensembles and ensures the calculation of elemental abundances for 
  concentric velocity intervals.
  \item For any component all lines corresponding to the same atomic or 
  ionic species are modelled by Doppler profiles of the same 
  broadening velocity and column density. With respect to their 
  broadening velocities all \ion{Cr}{ii}, \ion{Mn}{ii}, \ion{Fe}{ii}, 
  \ion{Ni}{ii}, and \ion{Zn}{ii} lines are modelled as if corresponding 
  to the same ionic species. The same treatment is applied to 
  \ion{Al}{ii} and \ion{Al}{iii} lines.
  \item Asymmetric lines like unresolved blends are modelled by a 
  single component if the individual components of the blend are separated
  by less than the full width at half maximum of the instrumental profile.
  \item Single absorption features with column density 
  $\log N < 13.8 - \log\lambda f$ 
  are ignored.
\end{enumerate}

All metal line ensembles are decomposed simultaneously while the local 
background continuum is approximated by an optimized linear combination 
of Legendre polynomials extending to the nearest absorption-free regions. 
The metal line recordings of STIS are too noisy and too contaminated with 
Lyman forest lines to be considered in the decomposition. The STIS 
Echelle order showing the sub-damped profile is decomposed using 
pseudo-Voigt profiles \citep{IdaAT_2000}. The profile is well defined by 
its Lorentzian part and can be distinguished from even a curved 
background continuum due to its symmetry. Note that the blue damping wing 
is blended with further \ion{H}{i} absorption which, however, does provide
only little information since due to saturation the solution is ambiguous.
For convenience, the blended absorption is
modelled by two components of the same width, and central wavelengths in
accord with the strongest associated metal lines (Fig.~\ref{fg:dla}). 

\subsection{Apparent optical depth analysis}

The apparent optical depth method is only applied to the weaker 
transitions of a given atomic or ionic species to avoid narrow 
saturation. The spectra are normalized using the optimized continuum 
approximation obtained from the line profile decomposition. The 
normalized flux is averaged using a moving window of 10~km\,s$^{-1}$. Low 
apparent optical depths $\tau_\mathrm{a}\le0.01$ are clipped.

\section{Results and discussion} 

In this section we present the optimized profile decomposition and 
examine the ionization, chemical composition and dust content, kinematic 
structure, and physical conditions in the absorbing medium.

\subsection{Profile decomposition}

The optimized decomposition of the sub-damped Ly~$\alpha$ profile and 
associated metal lines is depicted in Figs.~\ref{fg:dla}-\ref{fg:caii}. 
The corresponding line parameters are listed in Tables~\ref{tb:dla} 
and~\ref{tb:short}. Since only the metal line components 20-31 are 
associated with the sub-damped profile, components 1-19 and 20-31 are 
termed peripheral and main components, respectively. Note that components
23 and 24 correspond to the neutral carbon and H$_2$-bearing 
components considered by \citet{QuastBR_2002} and \citet{ReimersBQL_2003}.

\begin{table} \centering
\caption[]{Optimized decomposition of the sub-damped profile shown in 
Fig.~\ref{fg:dla}. The listed numbers correspond to a straight continuum, 
for a curved continuum the errors increase. The origin of the radial 
velocity (RV) scale corresponds to the redshift $z=1.15080$.
Note that the non-damped absorption does provide only little information.
The line parameters are listed for completeness only}
\begin{tabular}{@{}llll@{}}
\hline
\hline
Transition\rule[-5pt]{0pt}{15pt}
& $\phantom{-}\mathrm{RV}$ (km\,s\ensuremath{^{-1}})
& $b$ (km\,s\ensuremath{^{-1}})
& $\log N$ (cm\ensuremath{^{-2}})
\\
\hline
\ion{H}{i} 1216\rule{0pt}{10pt}
& $-3.19 \pm 11.14$
& $75.17 \pm 5.84$
& $19.89 \pm 0.03$
\\
\ion{H}{i} 1216
& $-236.14$
& $82.74 \pm 7.23$
& $16.84 \pm 0.92$
\\
\ion{H}{i} 1216
& $-533.51$
& $82.74 \pm 7.23$
& $14.98 \pm 0.13$
\\
\hline
\end{tabular}
\label{tb:dla}
\end{table}

\begin{table*} \centering
\caption[]{Optimized decomposition of the metal lines shown in 
Figs.~\ref{fg:metals} and \ref{fg:closeup}. This list is abridged for 
convenience, the complete table is available in the Online Material}
\begin{tabular}{@{}llllll@{}}
\hline
\hline
No.\rule[-5pt]{0pt}{15pt}
& \multicolumn{2}{l}{Transitions}
& $\phantom{-}\mathrm{RV}$ (km\,s\ensuremath{^{-1}})
& $b$ (km\,s\ensuremath{^{-1}})
& $\log N$ (cm\ensuremath{^{-2}})
\\
\hline
23\rule{0pt}{10pt}
& \ion{Mg}{i}
& 2026, 2853
& $-1.61 \pm 0.03$
& $2.16 \pm 0.05$
& $12.41 \pm 0.03$
\\
23
& \ion{Al}{ii}
& 1671
& $-1.61$
& $3.05$
& $13.86 \pm 0.17$
\\
23
& \ion{Al}{iii}
& 1855, 1863
& $-1.61$
& $3.05$
& $11.50 \pm 0.06$
\\
23
& \ion{Si}{i}
& 2515
& $-1.61 \pm 0.03$
& $2.17 \pm 2.10$
& $11.34 \pm 0.08$
\\
23
& \ion{Si}{ii}
& 1527, 1808
& $-1.61 \pm 0.03$
& $3.05 \pm 0.09$
& $14.28 \pm 0.03$
\\
23
& \ion{S}{i}
& 1807
& $-1.61 \pm 0.03$
& $1.56 \pm 1.12$
& $12.18 \pm 0.05$
\\
23
& \ion{Ca}{ii}
& 3935
& $-1.61 \pm 0.03$
& $2.15 \pm 0.10$
& $12.12 \pm 0.02$
\\
23
& \ion{Cr}{ii}
& 2056, 2062
& $-1.61 \pm 0.03$
& $2.97 \pm 0.05$
& $11.84 \pm 0.06$
\\
23
& \ion{Mn}{ii}
& 2577, 2594, 2606
& $-1.61 \pm 0.03$
& $2.97 \pm 0.05$
& $11.51 \pm 0.03$
\\
23
& \ion{Fe}{i}
& 2484, 2524
& $-1.61 \pm 0.03$
& $0.48 \pm 0.18$
& $11.30 \pm 0.06$
\\
23
& \ion{Fe}{ii}
& 1608, 2344, 2374, 2383, 2587, 2600
& $-1.61 \pm 0.03$
& $2.97 \pm 0.05$
& $13.51 \pm 0.01$
\\
23
& \ion{Ni}{ii}
& 1710, 1742, 1752
& $-1.61 \pm 0.03$
& $2.97 \pm 0.05$
& $12.48 \pm 0.04$
\\
23
& \ion{Zn}{ii}
& 2026, 2063
& $-1.61 \pm 0.03$
& $2.97 \pm 0.05$
& $11.79 \pm 0.02$
\\
\\
24
& \ion{Mg}{i}
& 2026, 2853
& $\phantom{-}7.63 \pm 0.06$
& $4.09 \pm 0.12$
& $11.90 \pm 0.01$
\\
24
& \ion{Al}{ii}
& 1671
& $\phantom{-}7.63$
& $5.36$
& $12.82 \pm 0.04$
\\
24
& \ion{Al}{iii}
& 1855, 1863
& $\phantom{-}7.63$
& $5.36$
& $11.72 \pm 0.05$
\\
24
& \ion{Si}{ii}
& 1527, 1808
& $\phantom{-}7.63 \pm 0.06$
& $5.36 \pm 0.25$
& $14.16 \pm 0.04$
\\
24
& \ion{Ca}{ii}
& 3935
& $\phantom{-}7.63 \pm 0.06$
& $3.17 \pm 0.17$
& $11.78 \pm 0.01$
\\
24
& \ion{Cr}{ii}
& 2056, 2062
& $\phantom{-}7.63 \pm 0.06$
& $4.92 \pm 0.13$
& $11.61 \pm 0.14$
\\
24
& \ion{Mn}{ii}
& 2577, 2594, 2606
& $\phantom{-}7.63 \pm 0.06$
& $4.92 \pm 0.13$
& $11.36 \pm 0.03$
\\
24
& \ion{Fe}{ii}
& 1608, 2344, 2374, 2383, 2587, 2600
& $\phantom{-}7.63 \pm 0.06$
& $4.92 \pm 0.13$
& $13.36 \pm 0.01$
\\
24
& \ion{Ni}{ii}
& 1710, 1742, 1752
& $\phantom{-}7.63 \pm 0.06$
& $4.92 \pm 0.13$
& $12.33 \pm 0.06$
\\
24
& \ion{Zn}{ii}
& 2026, 2063
& $\phantom{-}7.63 \pm 0.06$
& $4.92 \pm 0.13$
& $11.59 \pm 0.03$
\\
\\
28
& \ion{Mg}{i}
& 2026, 2853
& $\phantom{-}72.54 \pm 0.06$
& $6.03 \pm 0.15$
& $11.77 \pm 0.01$
\\
28
& \ion{Al}{ii}
& 1671
& $\phantom{-}72.54$
& $6.14 \pm 0.15$
& $12.57 \pm 0.02$
\\
28
& \ion{Al}{iii}
& 1855, 1863
& $\phantom{-}72.54$
& $6.14 \pm 0.15$
& $11.17 \pm 0.13$
\\
28
& \ion{Si}{ii}
& 1527, 1808
& $\phantom{-}72.54 \pm 0.06$
& $6.45 \pm 0.18$
& $13.75 \pm 0.02$
\\
28
& \ion{Ca}{ii}
& 3935
& $\phantom{-}72.54 \pm 0.06$
& $5.52 \pm 0.13$
& $12.08 \pm 0.01$
\\
28
& \ion{Cr}{ii}
& 2056, 2062
& $\phantom{-}72.54 \pm 0.06$
& $5.81 \pm 0.09$
& $11.92 \pm 0.06$
\\
28
& \ion{Mn}{ii}
& 2577, 2594, 2606
& $\phantom{-}72.54 \pm 0.06$
& $5.81 \pm 0.09$
& $11.34 \pm 0.03$
\\
28
& \ion{Fe}{ii}
& 1608, 2344, 2374, 2383, 2587, 2600
& $\phantom{-}72.54 \pm 0.06$
& $5.81 \pm 0.09$
& $13.51 \pm 0.01$
\\
28
& \ion{Ni}{ii}
& 1710, 1742, 1752
& $\phantom{-}72.54 \pm 0.06$
& $5.81 \pm 0.09$
& $12.32 \pm 0.06$
\\
28
& \ion{Zn}{ii}
& 2026, 2063
& $\phantom{-}72.54 \pm 0.06$
& $5.81 \pm 0.09$
& $10.72 \pm 0.11$
\\
\hline
\end{tabular}
\label{tb:short}
\end{table*}

\begin{figure*}
\centering
\includegraphics{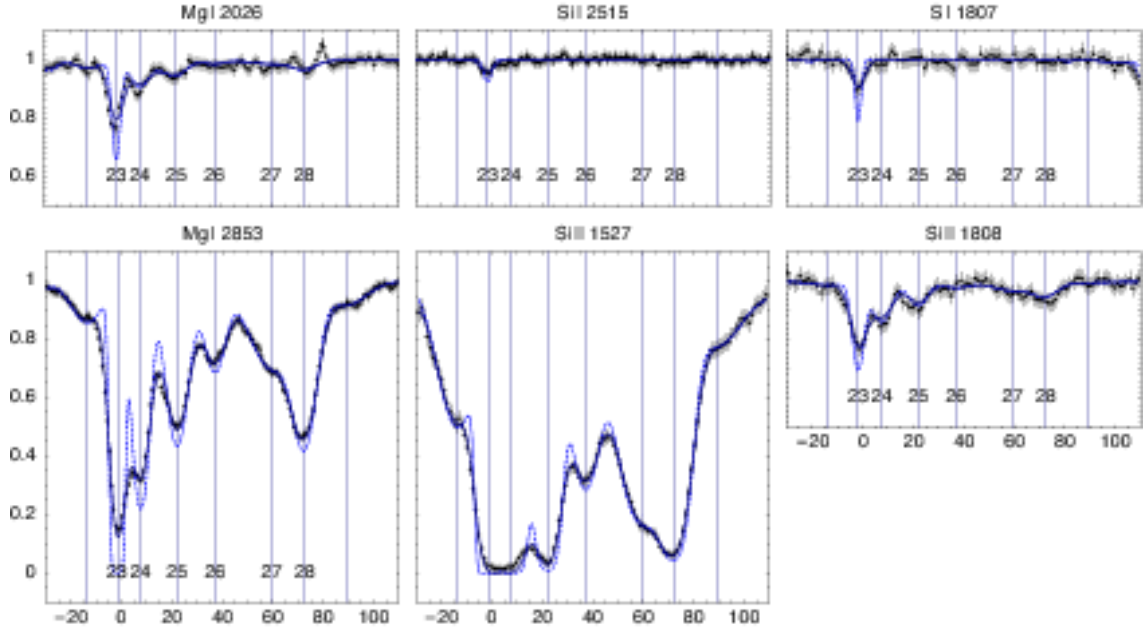}
\caption[]{Close-up of associated metal lines. Individual 
components are labeled by numbers 23-28}
\label{fg:closeup}
\end{figure*}

\addtocounter{figure}{-1}
\begin{figure*}
\centering
\includegraphics{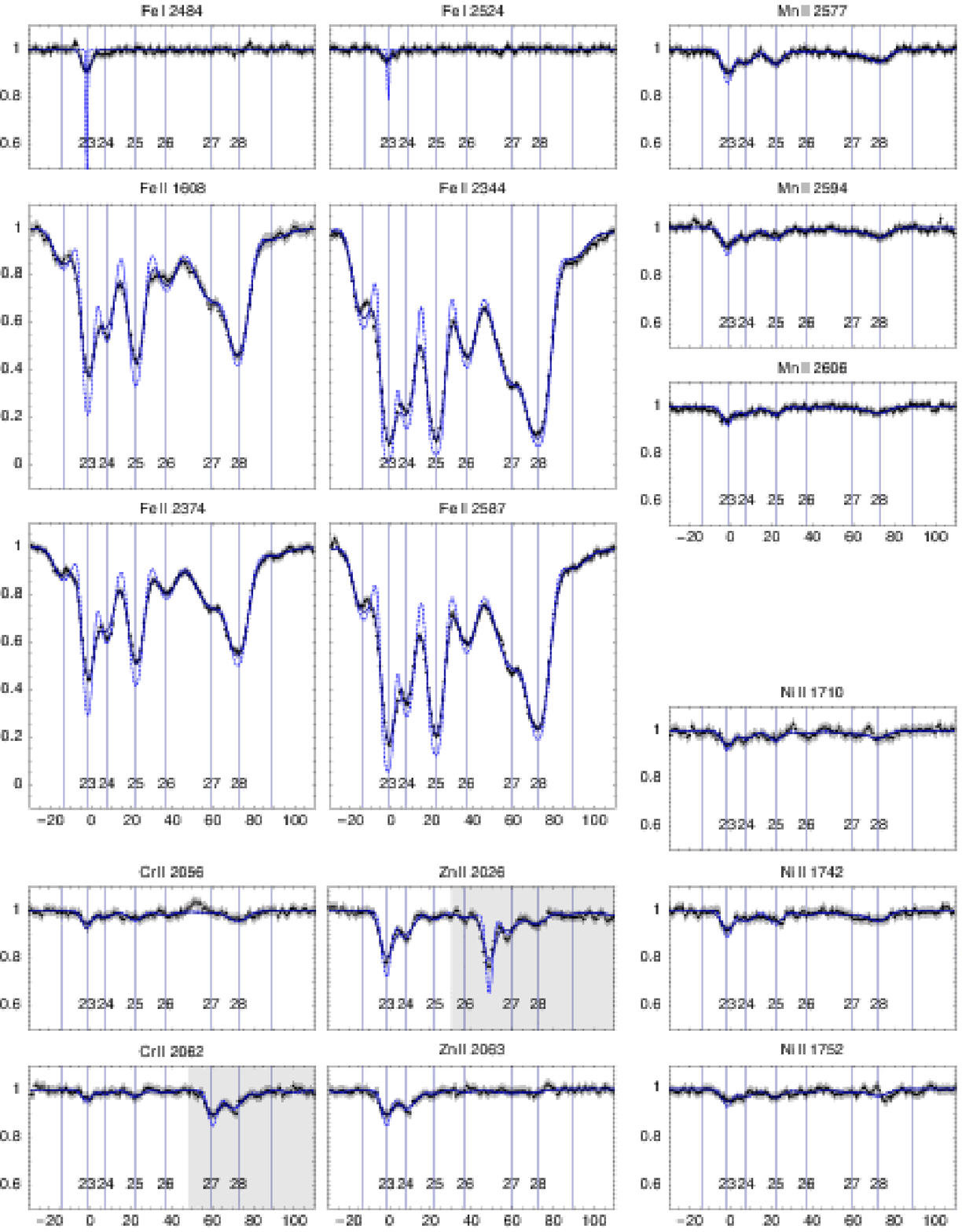}
\caption[]{continued. The shadings mark the main components of the 
\ion{Mg}{i}~$\lambda2026$ and \ion{Zn}{ii}~$\lambda2063$ lines.
Note that the detection of \ion{Fe}{i} and \ion{Si}{i} has been
reported earlier by \citet{QuastRS_2004}}
\end{figure*}

\begin{figure*} 
\centering
\includegraphics{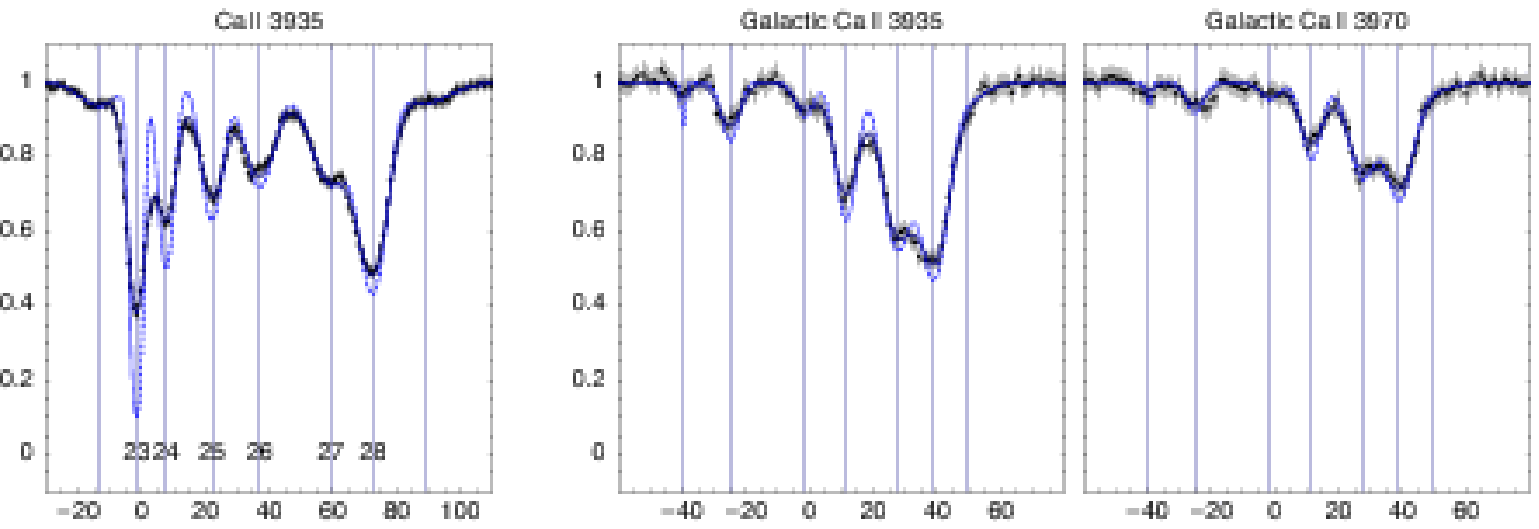}
\caption[]{Associated \ion{Ca}{ii} lines (left panel) compared with
interstellar \ion{Ca}{ii} absorption in the Galactic halo along the
same line-of-sight}
\label{fg:caii}
\end{figure*}

\subsubsection{Peripheral components 1-19}

The decomposition of the peripheral components is defined by the 
structure of \ion{Mg}{i}, \ion{Mg}{ii}, \ion{Si}{ii}, and \ion{Fe}{ii} 
lines. Part of the \ion{Si}{ii} profile is ignored due to contamination 
by Lyman forest lines. The weakest components with $10^{11}$ metal ions 
cm$^{-2}$ are indicated by the \ion{Mg}{ii} lines, whereas the components 
exceeding $10^{13}$ ions cm$^{-2}$ are saturated for \ion{Mg}{ii}, but 
well defined for \ion{Fe}{ii}. Nonetheless, the decomposition is 
uncertain in detail, since components 3, 16, and 17 possibly are 
unresolved blends. The ambiguities, however, do not affect the chemical 
abundance analysis.

For some components the \ion{Fe}{ii} lines are narrower than those 
corresponding to lighter elements. 
The constraints on the decomposition of \ion{Fe}{ii}, however, are much
stronger than those on the rest of lines. The evidence of thermal
broadening is therefore not conclusive.

\subsubsection{Main components 20-31}

The decomposition of the main components is well constrained by the 
structure of \ion{Mg}{i}, \ion{Si}{ii}, \ion{Ca}{ii} and \ion{Fe}{ii} 
lines. The \ion{Mg}{ii} lines are saturated, with column densities 
exceeding $2\times10^{13}$ ions cm$^{-2}$. The \ion{Al}{ii} line is 
ignored because it is blended with a lower-redshift \ion{Mg}{ii} system 
at $z=0.28$.\footnote{We cannot confirm the detection of \ion{Mn}{ii} 
lines associated with this DLA system candidate \citep{VargaRTBB_2000}.} 
Though the blended line ensemble can be disentangled, the optimized 
column densities calculated for components~23 and~24 are not reliable 
since the superimposed narrow \ion{Mg}{ii} absorption is saturated. The 
\ion{Si}{ii}~$\lambda1527$ line is saturated for components~23 and~24, 
but the optically thin \ion{Si}{ii}~$\lambda1808$ absorption compensates 
the lack of information. The red part of the \ion{Zn}{ii}~$\lambda2026$ 
line is blended with the blue part of \ion{Mg}{i}~$\lambda2026$, but due 
to the distinct \ion{Mg}{i}~$\lambda2853$ absorption both ensembles can 
be restored. Similarly, the red part of the \ion{Cr}{ii}~$\lambda2062$ 
line is blended with the blue part of \ion{Zn}{ii}~$2063$, but 
\ion{Cr}{ii}~$\lambda2056$ and the blue part of 
\ion{Zn}{ii}~$\lambda2026$ are unperturbed.

For the H$_2$-bearing components~23 and~24 the broadening velocity is 
correlated with the ionization potential of the absorbing species as if 
the ionizing radiation was spatially fluctuating. The lines corresponding 
to species with lower first ionization potential than hydrogen 
(\ion{C}{i}, \ion{Mg}{i} and \ion{Ca}{ii}) are systematically less 
broadened than the \ion{Si}{ii} and \ion{Fe}{ii} lines, indicating 
different spatial origins. This systematic difference is well known from 
the study of Galactic molecular gas 
\citep[Fig.~2]{SpitzerL_JenkinsE_1975}. Indeed, the detection of 
\ion{Si}{i}, \ion{S}{i}, and \ion{Fe}{i} lines with low broadening 
velocities (Table~\ref{tb:short}) may indicate an embedded layer of cold 
neutral gas. 
Note that the \ion{Fe}{i} absorption lines toward HE~0001--2340 that have
recently been detected by \citet{OdoricoV_2007} also show an extremely low
broadening parameter.

\subsection{Ionization}

For most elements, the singly ionized state predominates in the neutral 
interstellar medium because the first ionization potential is lower, 
whereas the second is higher than the hydrogen ionization threshold. 
Exceptions to this rule are N, O, and Ar, where the first ionization 
potential exceeds the threshold, and Ca, where even the second ionization 
potential is lower. Since for interstellar abundance studies the total 
amount of an element is usually assumed to be equal to the amount
existing in the predominant stage of ionization, substantial errors are
made if the absorbing medium is a mixture of \ion{H}{i} and \ion{H}{ii}
regions.

\subsubsection{Peripheral components}

Immediate evidence of ionized gas is provided by the detection of 
\ion{C}{iv} lines (Fig.~\ref{fg:metals}) and the presence of 
\ion{Si}{iii} and \ion{Si}{iv} absorption (see Fig.~\ref{fg:stis} 
provided in the Online Material). Apparent column densities of up to
$10^{13}$~cm$^{-2}$ are found for all high-ions. Except for the broad 
\ion{C}{iv} lines, the velocity structures of low- and high-ion profiles 
are similar, suggesting a common spatial origin. In particular for 
components 2-4 the apparent column densities of different Si ions 
compare, which indicates an \ion{H}{ii} region. The apparent optical 
depths of \ion{Si}{iii} and \ion{Si}{iv} decreases for components 7-16,
whereas the optical depth of \ion{Si}{ii} peaks at component 16. The
optically thick absorption between components 19 and 20 of the
\ion{Si}{iii} profile indicates ionization, but \ion{Si}{iii} absorption
for this velocity interval is not confirmed by the rest of metal lines.

\addtocounter{figure}{1}
\begin{figure}
\centering
\includegraphics{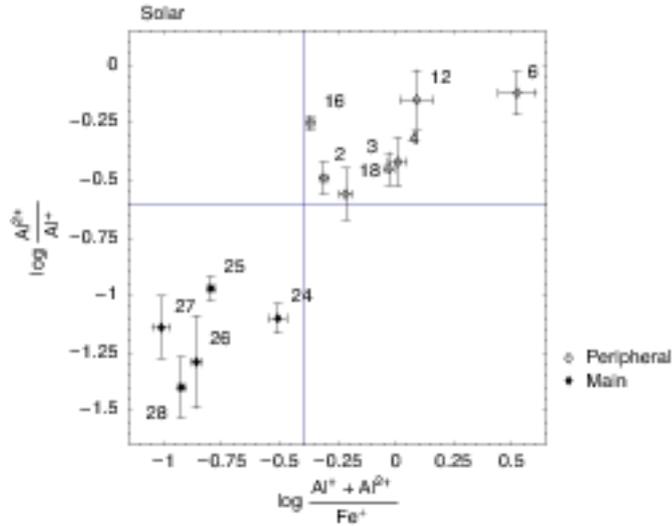}
\caption[]{Empirical diagram of ionization vs apparent abundance of Al
relative to Fe. Note that the peripheral and main components are well
separated}
\label{fg:twocolor}
\end{figure}

Firm evidence of ionized gas is provided by the column density ratios 
of \ion{Al}{ii}, \ion{Al}{iii}, and \ion{Fe}{ii} lines. For components 2-4,
6, 12, 16, and 18 the apparent abundance of Al relative to Fe is a factor 
of 4-30 higher than expected for a neutral medium with solar chemical 
composition (Fig.~\ref{fg:twocolor}). The apparent enrichment of Al 
cannot be explained by the presence of dust because the expected 
depletion of Al into grains is typically an order of magnitude higher 
than that of Fe \citep{SpitzerL_JenkinsE_1975}. On the other hand, Al is 
produced with $\alpha$-elements which are known to experience a 
nucleosynthetic history different from that of Fe. For instance, in 
Galactic thick disk stars the abundance ratio of Al to Fe is found to be 
enhanced by a factor of 2-4 relative to the solar value 
\citep{ProchaskaNC_2000}. Nonetheless, the apparent enrichment of Al is 
correlated with the column density ratio of \ion{Al}{iii} to \ion{Al}{ii} 
lines, indicating ionization rather than nucleosynthetic enrichment 
(Fig.~\ref{fg:twocolor}). 

\begin{figure*}
\centering
\includegraphics{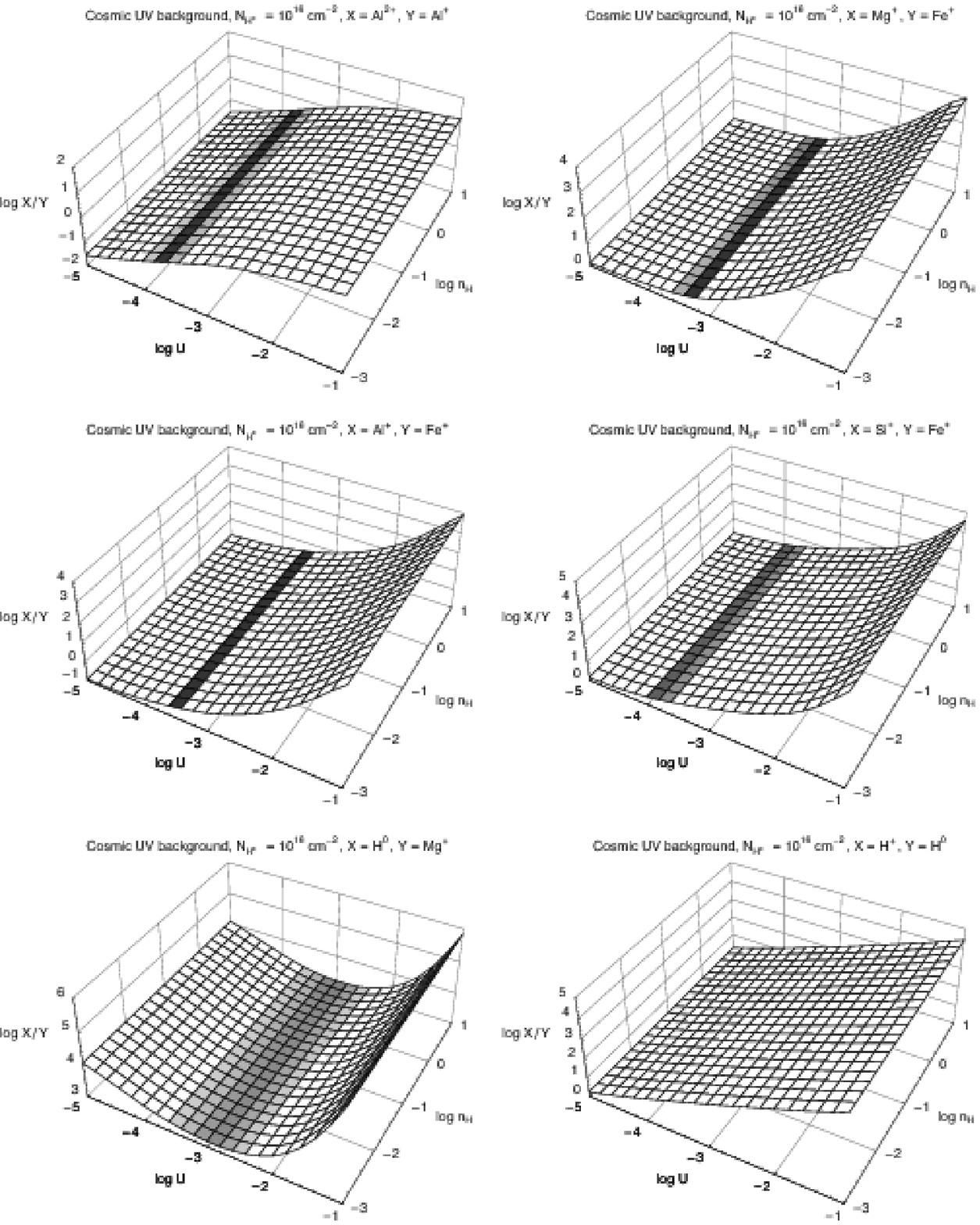}
\caption[]{Photoionization simulations considering a plane-parallel slab 
of gas with column density of $10^{16}$ hydrogen atoms cm$^{-2}$ and 
solar chemical composition. The abscissas mark the ionization parameter 
$U$ and the total hydrogen density $n_\mathrm{H}$. Shaded cells indicate 
consistency with the observed column density ratios for component~2, 
unshaded cells mark contradiction at more than 99.7 percent significance. 
For the medium corresponding to component~2 where $U>10^{-4}$ the 
predicted degree of ionization exceeds 90 percent. Note that the observed 
column density ratios do pretend an enrichment of $\alpha$-elements when 
ionization effects are ignored}
\label{fg:cloudy16}
\end{figure*}

Further confidence is provided by photoionization simulations. For the 
calculations we consider a plane-parallel slab of gas that is irradiated 
by the cosmic UV background of QSOs and galaxies 
\citep{MadauHR_1999}.\footnote{The photoionization simulations have been 
carried out with version 05.07 of Cloudy, last described by 
\citet{FerlandKV_1998}. The cosmic UV background at redshift 1.15 has been 
calculated using lookup tables provided by F.~Haardt.} We further assume 
a column density of $10^{16}$ hydrogen atoms cm$^{-2}$ and solar chemical 
composition. The photoionization models are defined by the total hydrogen 
density $n_\mathrm{H}$ and the dimensionless ionization parameter
\begin{equation}
U = \frac{n_\gamma}{n_\mathrm{H}},
\end{equation}
where
\begin{equation}
n_\gamma =
\frac{4\pi}{c}\int_{\nu_\mathrm{L}}^\infty\frac{J(\nu)}{h\nu}\,\mathrm{d}\nu
\end{equation}
is the number density of hydrogen-ionizing photons striking the 
illuminated face of the slab. For instance, for a total hydrogen density 
of $n_\mathrm{H}=10^{-1}$ particles cm$^{-3}$ the cosmic UV background 
with $4\pi J_{\nu_\mathrm{L}}=8.6\times10^{-21}$~erg\,cm$^{-2}$ 
corresponds to an ionization parameter of $U=2\times 10^{-4}$. 
Figure~\ref{fg:cloudy16} demonstrates that the observed column density 
ratios are well reproduced, if $U$ falls between $10^{-4}$ and $10^{-3}$. 
The only exception is the ratio of \ion{Mg}{ii} to \ion{Mg}{i}, which is
overpredicted by a factor of two. 
The degree of ionization for this range of $U$ is higher than 90 percent. 
Similar results are obtained for column densities of $10^{17}$ and 
$10^{18}$ hydrogen atoms cm$^{-2}$ and column density ratios 
corresponding to component~16.

\begin{figure*}
\centering
\includegraphics{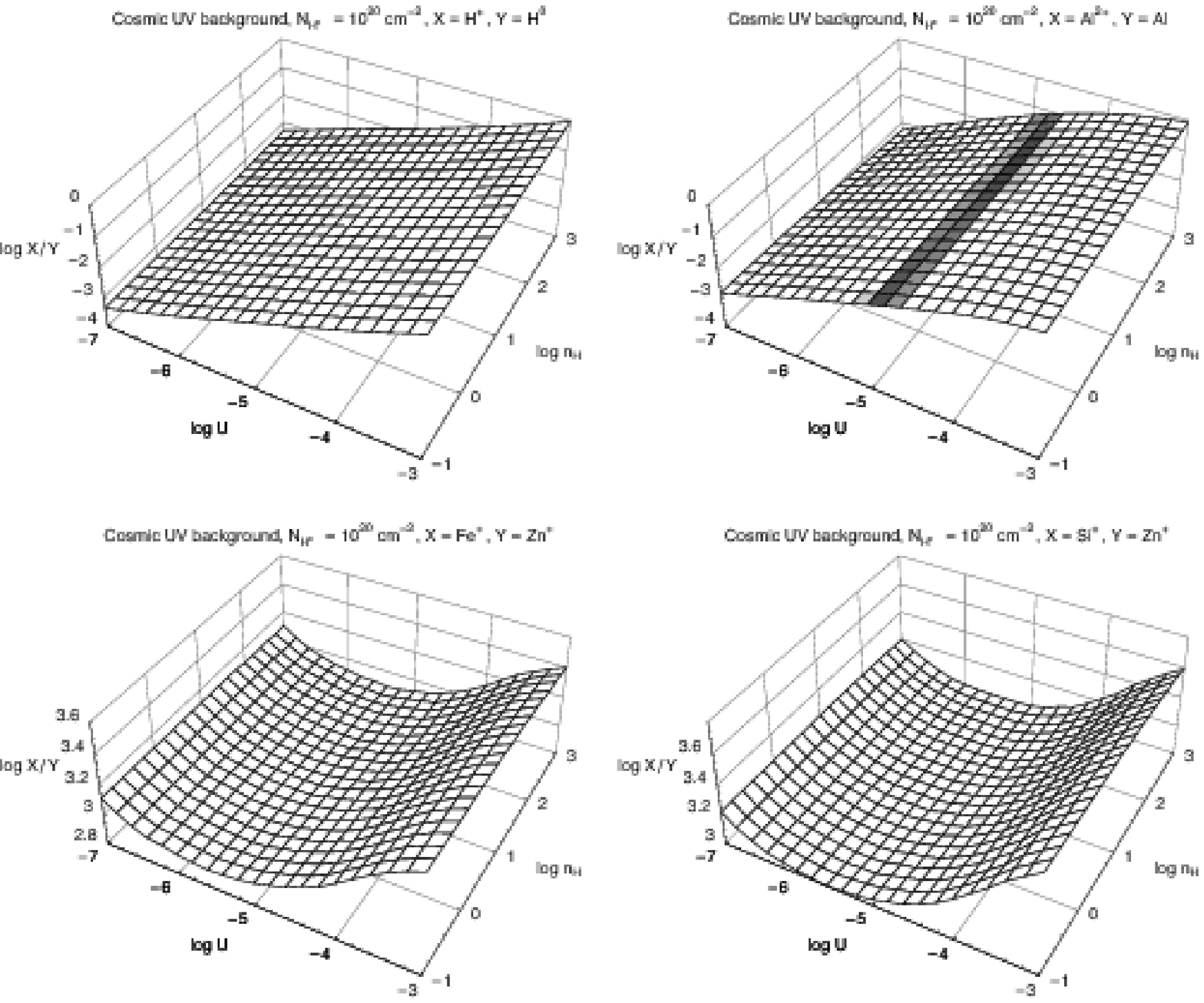}
\caption[]{The same as Fig.~\ref{fg:cloudy16} but a plane parallel slab
with column density of $10^{20}$ hydrogen atoms cm$^{-2}$ is considered.
\emph{Top:}
For a total hydrogen density of $10^{-1}$ particles cm$^{-3}$ the cosmic 
UV background corresponds to an ionization parameter of $U=2\times 
10^{-4}$. For denser regions like molecular gas the same ionizing 
background corresponds to $U<10^{-4}$, implying that at most 10~percent 
of the hydrogen is ionized. This argument is confirmed by the observed
column density ratio of Al ions (illustrated by the shading for component
24, but also see Fig.~\ref{fg:twocolor}).
\emph{Bottom:}
The minimum of the predicted ratio of Fe and Zn ion abundances 
corresponds to the case where both elements almost completely reside in 
the first ionization stage and the abundance ratio of ions is almost 
equal to the (solar) abundance ratio of elements. Note that ionization 
effects cannot pretend a depletion of Fe relative to Zn. The situation 
for Si and Zn is similar}
\label{fg:cloudy20}
\end{figure*}

In summary, there is conclusive evidence that the peripheral metal line 
components are formed within \ion{H}{ii} regions.
An empirical method to identify a \ion{H}{ii} region when individual
\ion{H}{i} components are not detected is provided by
Figure~\ref{fg:twocolor}. 

\subsubsection{Main components}

The presence of ionized gas within the main absorber is not as evident as 
for the periphery. The \ion{Si}{iii} profile is optically thick, 
suggesting an apparent column density possibly exceeding $10^{13}$ ions 
cm$^{-2}$ for components 20-28, but most of the absorption is due to 
Lyman forest lines (see the preceding subsection). While the stronger 
\ion{Si}{iv} profile is blended into the Lyman forest, the weaker 
\ion{Si}{iv} profile may confirm the \ion{Si}{iii} absorption for 
components 27-31. Conclusive evidence of ionized gas is provided by the 
\ion{C}{iv} and \ion{Al}{iii} lines. The velocity structure of the 
\ion{C}{iv} line, with an apparent column density of up to 
$2\times10^{12}$ ions cm$^{-2}$, is weak and without noticeable 
substructure in the domain of components 22-26, indicating that ionized 
and neutral gas are not intermixed. In contrast, the \ion{Al}{iii} 
profiles indicate a homogenous distribution of high- and low-ions for 
components 23-25. The column density ratio of \ion{Al}{iii} to 
\ion{Al}{ii} lines is less than $1/10$ for all components 
(Fig.~\ref{fg:twocolor}).

According to simple photoionization calculations an ionization of the 
molecular regions due to the cosmic UV background is ruled out 
(Fig.~\ref{fg:cloudy20}). However, some fractional ionization due to 
stellar sources is conceivable. The comparison with photoionization 
simulations considering both interstellar radiation and the formation of 
molecules and dust requires an accurate recording of H$_2$ lines that is 
not available. 
Our simple attempts assuming Galactic environmental 
conditions have yielded inconsistent results, which reproduce the column
density ratios of \ion{Fe}{i} to \ion{Fe}{ii} and \ion{Si}{i} to
\ion{Si}{ii} as well as the relative population of neutral carbon
fine-structure levels, but both overpredict the amount of molecular
hydrogen and underpredict the strength of the \ion{Ca}{ii} absorption by
more than one order of magnitude. Our calculations hence suggest that the
radiation field and the density structure of the absorber are more
complex.
In fact, for component 23 the neutral species \ion{Fe}{i}, \ion{Si}{i} and
\ion{S}{i} have much lower broadening parameters than \ion{Fe}{ii} and
\ion{Si}{ii}, which indicates that the different ionization stages
are not formed in the same region. Photoionization calculations without
modelling the density structure are therefore meaningless. Similar
conclusions have been drawn by \citet{OdoricoV_2007} who failed to reproduce
the observed \ion{Mg}{i} to \ion{Mg}{ii}, \ion{Fe}{i} to \ion{Fe}{ii} and
\ion{Ca}{i} to \ion{Ca}{ii} ratios in a metal line system toward HE 0001-2340. 

\subsection{Metal abundances and dust depletion}
\label{sc:chem}

Besides ionization and nucleosynthetic effects, the chemical composition 
analysis of interstellar environments is hampered by dust grains removing 
an unknown amount of atoms from the gaseous phase 
\citep{SpitzerL_JenkinsE_1975,SavageB_SembachK_1996}. The accepted 
procedure to unravel these effects is to compare the abundance of 
refractory and volatile elements $X,Y$ for which the photospheric 
abundance ratio $(X/Y)$ is constant in stars over a wide range of 
metallicities. In that case an observed deviation from the stellar ratios 
is unlikely to have a nucleosynthetic origin. Even though the existence 
of a stellar proxy for interstellar abundances is questionable 
\citep{SofiaU_MeyerD_2001} the Sun is used as a standard of reference for 
the total, i.e. gas plus dust, interstellar composition. For given 
observed column densities $N_X$ and $N_Y$ the relative abundance of 
elements $X$ and $Y$ is expressed as
\begin{equation}
[X/Y] = \log(N_X/N_Y) - \log(X/Y)_{\sun}.
\end{equation}
Element abundances relative to hydrogen are termed absolute abundances. 
âFor chemical composition analysis, we use the metal abundances in 
meteorites \citep{AndersE_GrevesseN_1989} as a standard of reference.

\subsubsection{Peripheral components}

Since the peripheral components are formed within \ion{H}{ii} regions, 
the element abundances cannot be determined directly. The observed ion 
abundances, however, are supersolar and photoionization calculations 
conform with the idea that both absolute and relative metal abundances 
are solar (Fig.~\ref{fg:cloudy16}).

\subsubsection{Main components}

For the main components photoionization calculations suggest that the 
absorbing material is predominantly neutral and ionization effects are 
negligible, i.e. all elements are accurately represented by the 
predominant ions. For the chemical composition analysis and the proper 
unravelling of dust depletion and nucleosynthetic effects the detection 
of volatile elements is essential. The only volatile element detected in 
the predominant ionization stage with accurate column density 
measurements for several main components is Zn. For other volatile 
elements like N and O also detected in the predominant ionization stage, 
the absorption is saturated and largely blended with Lyman forest lines 
(Fig.~\ref{fg:stis}). Even though the \ion{Zn}{ii} absorption is weak, 
the individual column densities are well defined because the positional 
and broadening parameters of the decomposed \ion{Zn}{ii} profiles are 
tied to those of the \ion{Fe}{ii} lines.

\begin{figure} \centering
\includegraphics{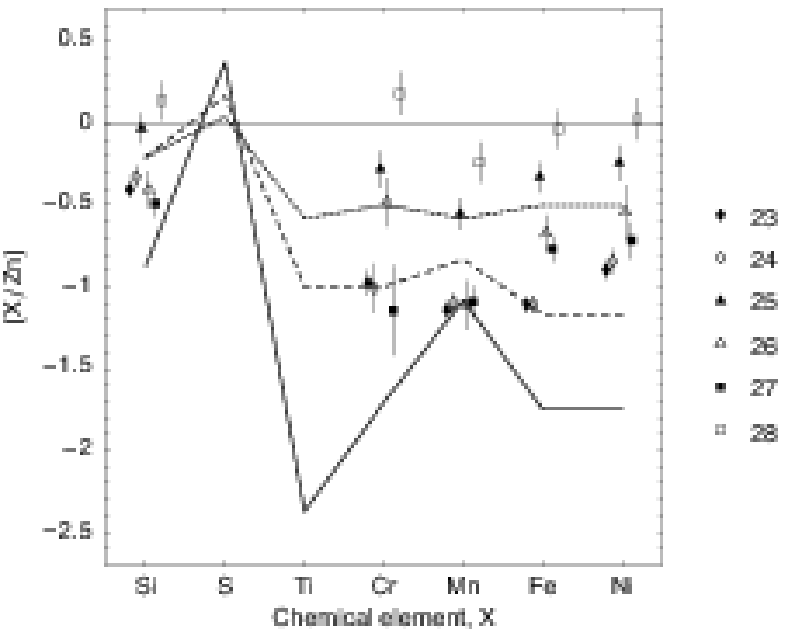}
\caption[]{Gas-phase abundance ratios (relative to solar ratios) for the 
main components 23-28 compared with those in the Galactic interstellar 
medium (dotted line: warm halo gas; dashed line: warm disk gas; solid 
line: cold disk gas; \citealp{WeltyLBHY_2001}). If the volatile element 
Zn is undepleted, the abundance ratios reflect the differential depletion 
of chemical elements into dust grains. Note that Cr, Mn, Fe, and Ni are 
strongly depleted for the H$_2$-bearing components 23 and 24, but
essentially undepleted for component 28}
\label{fg:gas}
\end{figure}

\paragraph{Gas-phase abundances}
The observed abundances of refractory elements relative to Zn for 
components 23-28 are illustrated in Fig.~\ref{fg:gas}. For components 23 
and 24 the underabundance of the iron group elements Cr, Mn, Fe and Ni 
relative to Zn is comparable to that in the Galactic warm disk,
similar to what has been found by \citet{RodriguezPALS_2006},
whereas the mild and even vanishing depletion for components 25 and 28 is not 
found in Galactic interstellar space 
\citep{SpitzerL_JenkinsE_1975,SavageB_SembachK_1996}. The relative 
abundances found for components 26 and 27 rather resemble those of the 
Galactic halo. If Zn is indeed undepleted and traces Fe as found by 
\citet{NissenCA_2004}, the pattern of relative abundances directly 
reflects the differential depletion of chemical elements into dust 
grains. This idea gains indirect support by the detection of H$_2$ 
absorption lines associated with components 23 and 24, since molecules 
are essentially formed on the surface of dust grains 
\citep{CasauxCT_2005,WilliamsD_2005}. 
For all components showing evidence 
of dust grains, the depletion of Si is stronger than in the Galactic warm 
disk, but weaker than in the cold disk. For all but the H$_2$-bearing 
components Mn is systematically underabundant when compared to the rest 
of iron group elements. The synthesis of Mn, however, is expected to be 
suppressed due to the nuclear odd-even effect. The average metallicity 
for components 23-28 is $[\mathrm{Zn}/\mathrm{H}]=-0.38\pm0.04$.

The apparent underabundance of Fe (and Si) relative to Zn cannot be the 
result of ionization effects caused by the cosmic UV background since 
these would pretend an enhanced abundance (Fig.~\ref{fg:cloudy20}). 
Therefore, the interpretation of the observed underabundance as evidence 
of depletion into dust grains cannot be questioned without admitting very 
unusual nucleosynthetic effects. On the other hand, if the observed 
underabundance of Fe (and Si) relative to Zn is the net result of dust 
depletion and ionization effects, the true depletion might be even 
stronger than illustrated in Fig.~\ref{fg:gas}.

In summary, there is decisive evidence of Galaxy-like differential 
depletion of elements into dust grains, with a significant gradient from 
component to component as if the sight line is intersecting different 
interstellar environments comparable to the Galactic disk and halo. 
Another such example may be the DLA system toward the gravitationally 
lensed QSO HE~0512--3329, where different element abundances are detected 
along two lines of sight \citep{LopezRG_2005}. Further examples are given 
by \citet{DessaugesPD_2006}.

\paragraph{Dust correction}
Based on the presumption that the chemical composition of dust is defined 
by the physical state and the chemical composition of the medium, 
\citet{VladiloG_2002a,VladiloG_2002b} has worked out an analytic relation 
between the dust depletion patterns for interstellar environments of 
different types. Though generalizing former approaches where the dust 
composition is assumed to be constant, his new approach still implies 
that the dust composition does not depend on the history of the medium, a 
condition that is not strictly satisfied since the formation of dust and 
ices involves irreversible processes \citep{VidaliRM_2005}.

For any interstellar environments $i,j$ and constant sensitivity of the 
chemical composition of dust to variations in the dust-to-metal ratio 
$\rho$ and the chemical composition of the medium, the fractions of the 
generic and reference elements $X,Y$ contained in dust grains are related
by
\begin{equation}
f_{X,j} = \rho^{\eta_X+1} 10^{[X/Y]_{\mathrm{m},j}\,(\epsilon_X-1)} 
f_{X,i}.
\label{eq:fraction}
\end{equation}
The subscript \lq m\rq\ indicates reference to all atoms in the medium,
i.e. in the gaseous and the solid phase. The exponents 
$\eta_X,\epsilon_X$ define the response of the relative abundance of $X$ 
in the solid phase to the variation of the fraction of $Y$ contained in 
dust, and the relative abundance of $X$ in the medium, respectively 
\citep{VladiloG_2002a,VladiloG_2002b}.

Equation~(\ref{eq:fraction}) is capable of reproducing the Galactic 
interstellar depletion patterns $\delta_X =\log(1-f_X)$ with a varying 
dust-to-metal ratio
\begin{equation}
\rho =
\frac{f_{Y,j}}{f_{Y,i}}
\label{eq:rho}
\end{equation}
and a single set of empirical constants $(f_{X,i},\eta_X)$.\footnote{For 
the Galaxy, the exponent $\epsilon_X$ is irrelevant since the 
interstellar element abundances are assumed to be solar.} The depletion 
patterns in the interstellar medium of the SMC can be reproduced with the 
same set of constants, if the relative element abundances are allowed to 
deviate from solar values. From the theoretical point of view, the 
dust-corrected element abundances of high-redshift DLA systems show 
better consistency with galactic chemical evolution models than the plain 
observations \citep{CaluraMV_2003}.

An explicit relation between observed and intrinsic absolute abundances 
is obtained by using Eq.~(\ref{eq:fraction}) to express the fraction of 
$X$ contained in the gaseous phase of medium $j$:
\begin{equation}
\frac{10^{[X/\mathrm{H}]_{\mathrm{g},j}}}{10^{[X/\mathrm{H}]_{\mathrm{m},j}}} 
= 1 - \rho^{\eta_X+1} 10^{[X/Y]_{\mathrm{m},j}\,(\epsilon_X-1)} f_{X,i},
\label{eq:absolute}
\end{equation}
where the subscript \lq g\rq\ indicates reference atoms in the gaseous 
phase. The parameters $f_{X,i}$ and $\eta_X$ can be calculated from 
Galactic interstellar depletion patterns 
\citep{VladiloG_2002a,VladiloG_2002b}, whereas the dust-to-gas ratio 
$\rho$ is an implicit function of the observed and intrinsic abundance 
ratios
\begin{equation}
\rho -
\frac{10^{[X/Y]_{\mathrm{m},j}\,\epsilon_X}f_{X,i}}{10^{[X/Y]_{\mathrm{g},j}}f_{Y,i}}
\rho^{\eta_X+1} + \frac{10^{[X/Y]_{\mathrm{m},j}} - 
10^{[X/Y]_{\mathrm{g},j}}}{ 10^{[X/Y]_{\mathrm{g},j}} f_{Y,i}} = 0,
\label{eq:implicitrho}
\end{equation}
which is obtained by dividing Eq.~(\ref{eq:absolute}) by the 
corresponding equation for $X=Y$. The intrinsic abundance ratio 
$[X/Y]_{\mathrm{m},j}$ and the exponent $\epsilon_X$ are unknown 
parameters. For the latter only two extreme cases are considered where 
the relative element abundances in the solid phase and the medium are 
mutually independent, $\epsilon_X=0$, or directly proportional, 
$\epsilon_X=1$. Superlinear response $\epsilon_X>1$ is ignored. For the 
intrinsic abundance ratio $[X/Y]_{\mathrm{m},j}$ an element $X=Z$ ideally 
tracing the reference element $Y$ is required.

Given the observed abundance ratios $10^{[X/Y]_{\mathrm{g},j}}$ and an 
educated guess of $[Z/Y]_{\mathrm{m},j}$, the dust-to-metal ratio $\rho$ 
is defined by Eq.(\ref{eq:implicitrho}). The rest of intrinsic abundance 
ratios $[X/Y]_{\mathrm{m},j}$ for elements $X\ne Z$ implicitly follows 
from
\begin{equation}
10^{[X/Y]_{\mathrm{m},j}} - \rho^{\eta_X+1} 
f_{X,i}\,10^{[X/Y]_{\mathrm{m},j}\,\epsilon_X} + (\rho 
f_{Y,i}-1)\,10^{[X/Y]_{\mathrm{g},j}} = 0.
\label{eq:implicitrel}
\end{equation}
The roots of Eqs.~(\ref{eq:implicitrho}, \ref{eq:implicitrel}) can be 
calculated with simple bisectioning and the statistical errors of all 
quantities involved in the calculation, i.e. observed column densities, 
meteoritic element abundances, $f_{X,i}$, $\eta_X$, can be propagated by 
means of Monte Carlo methods. The intrinsic absolute element abundances 
$[X/\mathrm{H}]_{\mathrm{m},j}$ follow from Eq.~(\ref{eq:absolute}). The 
dust-to-gas ratio is given by
\begin{equation}
\kappa =
\rho\,10^{[Y/\mathrm{H}]_{\mathrm{m},j}}.
\label{eq:kappa}
\end{equation}

For the corrective procedure we choose $Y=\mathrm{Fe}$ since the 
\ion{Fe}{ii} lines provide the most reliable column densities. The most 
suitable element for calculating the dust-to-metal ratio is 
$Z=\mathrm{Zn}$. The rest of elements with known parameters $f_{X,i}, 
\eta_X$ are either too affine to the solid state, or have an interstellar 
enrichment history too different from that of Fe to make an assumption 
about the intrinsic abundance ratio. If Zn proves a tracer of Si as 
suggested by \citet{WolfeGP_2005}, the combination $Y=\mathrm{Si}$ and 
$Z=\mathrm{Zn}$ will be an eligible option. For the present case, 
however, this option results in implausible dust-to-metal ratios 
throughout exceeding those in the Galactic cold disk. Following 
\citet{VladiloG_2004}, we consider six different model cases labeled with 
A0, A1, B0, B1, E0, and E1. The capital letter indicates the intrinsic 
abundance of Zn relative to Fe, assumed solar for cases A and B, and 
enhanced for case E, $[\mathrm{Zn}/\mathrm{Fe}]_{\mathrm{m},j}=0.1$. For 
cases A and E the fraction of Zn assumed to be contained in interstellar 
dust is $f_{\mathrm{Zn},i}=0.59$, for case B the assumed depletion is 
lower, $f_{\mathrm{Zn},i}=0.30$. The number to the right of the capital 
letter directly represents the value of the exponent $\epsilon_X$. Zinc 
might be no precise tracer of iron since its exact nucleosynthetic origin 
is unknown, but the average photospheric abundance ratios in Galactic 
disk and halo stars are approximately solar or slightly enhanced 
\citep{ChenNZ_2004,NissenCA_2004} and substantiate the model 
specifications.

\begin{figure} \centering
\includegraphics{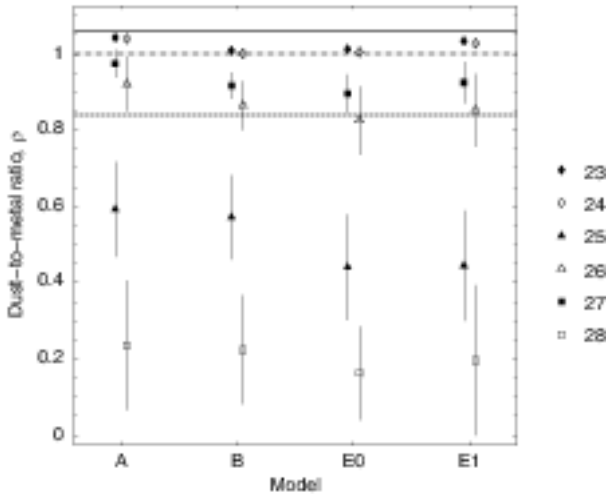}
\caption[]{Dust-to-metal ratio $\rho$ for the main components 23-28 
compared with values typical for the Galactic interstellar medium (dotted 
line: warm halo gas; dashed line: warm disk gas; solid line: cold disk 
gas; \citealp{VladiloG_2002b}). The results for models A and B do not 
depend on the exponent $\epsilon_X$}
\label{fg:rho}
\end{figure}

\begin{figure*} \centering
\includegraphics{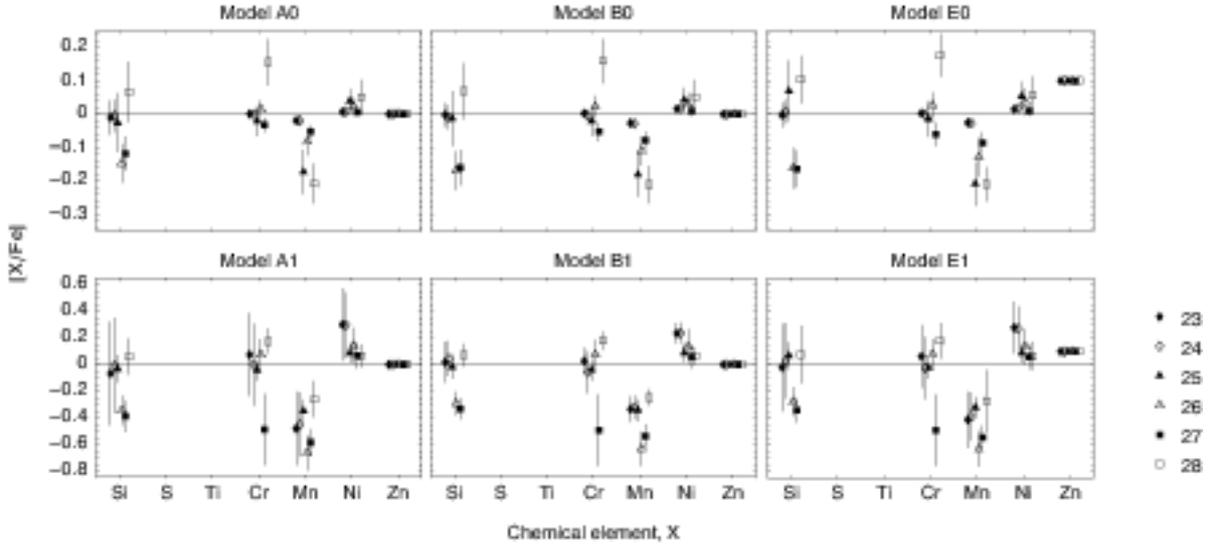}
\caption[]{Intrinsic abundance ratios (relative to solar ratios) for the 
main components 23-28}
\label{fg:intrinsic}
\end{figure*}

\paragraph{Dust-to-metal ratio}
The calculated dust-to-metal ratios and corresponding fractions of Fe 
contained in dust are presented in Fig.~\ref{fg:rho} and 
Table~\ref{tb:ffe}, respectively. If there is a one-to-one correspondence 
between the physical state of the interstellar environment and the 
dust-to-metal ratio, Fig.~\ref{fg:rho} illustrates the multiphase 
structure of the absorbing medium. Though such diversified structure is 
characteristic for the interstellar medium in the Galaxy and the 
Magellanic Clouds
\citep{SavageB_SembachK_1996,WeltyLBHY_1997,WeltyFSY_1999,WeltyLBHY_2001} 
it is usually not attributed to DLA systems \citep{ProchaskaJ_2003}. The 
dust-to-metal ratio for the H$_2$-bearing components 23 and 24 exceeds 
the ratio of the warm Galactic disk, for some models the ratio approaches 
the ratio found in the cold disk. For components 26 and 27 the 
dust-to-metal ratio compares to that in the intermediate warm Galactic 
disk and halo, whereas for the the rest of components the dust-to-metal 
ratio is typical of higher-redshift DLA systems \citep{VladiloG_2004}. 
These basic results conform with the observed depletion pattern 
(Fig.~\ref{fg:gas}) and are independent of the adopted model, in 
particular unaffected by the choice of 
$[\mathrm{Zn}/\mathrm{Fe}]_{\mathrm{m},j}$.

\paragraph{Intrinsic abundances and dust-to-gas ratio}
The calculated intrinsic abundance ratios are presented in 
Fig.~\ref{fg:intrinsic}. For all models the intrinsic abundance ratios 
conform with solar values, apart from three exceptions: 
\begin{enumerate}
  \item The mean intrinsic abundance of Ni relative to Fe is slightly
  enhanced. Similar offsets are also found for Galactic thick disk stars 
  \citep{ProchaskaNC_2000}.
  \item The intrinsic abundance of Mn is reduced. The reduction is most 
  distinct for the less dust-containing components 25-28 and for models 
  where the relative element abundances in dust and in the medium scale 
  directly, i.e. $\eta_X=1$. Similar underabundances, usually attributed 
  to the nuclear odd-even effect, are found for Galactic thick disk stars 
  \citep{ProchaskaNC_2000} as well as DLA systems 
  \citep{DessaugesPD_2006}.
  \item For the intermediate components 26 and 27 the intrinsic abundance 
  of Si relative to Fe is always subsolar, in marked contrast to the 
  expected nucleosynthetic enrichment of $\alpha$-elements and to element 
  abundances found for Galactic stars. For these components, the 
  dust-corrective procedure may have overestimated (underestimated) the 
  fraction of Fe (Si) contained in dust.
\end{enumerate}

\begin{table*} \centering
\caption{Fractions of Fe contained in dust, $f_{\mathrm{Fe},j}$, for the 
main components 23-28 and different models. The results for models A and 
B do not depend on the exponent~$\epsilon_X$}
\begin{tabular}{@{}llllll@{}}
\hline
\hline
No.\rule[-5pt]{0pt}{15pt}
& A
& B
& E0
& E1
& Mean
\\
\hline
23\rule{0pt}{10pt}
& $0.98\pm0.01$
& $0.95\pm0.01$
& $0.95\pm0.01$
& $0.97\pm0.01$
& $0.96\pm0.02$
\\
24
& $0.98\pm0.01$
& $0.94\pm0.01$
& $0.94\pm0.01$
& $0.96\pm0.01$
& $0.96\pm0.02$
\\
25
& $0.56\pm0.12$
& $0.54\pm0.10$
& $0.41\pm0.13$
& $0.42\pm0.14$
& $0.48\pm0.08$
\\
26
& $0.86\pm0.07$
& $0.81\pm0.06$
& $0.77\pm0.08$
& $0.80\pm0.09$
& $0.81\pm0.04$
\\
27
& $0.92\pm0.03$
& $0.86\pm0.03$
& $0.84\pm0.05$
& $0.87\pm0.05$
& $0.87\pm0.03$
\\
28
& $0.22\pm0.16$
& $0.21\pm0.13$
& $0.15\pm0.11$
& $0.18\pm0.18$
& $0.19\pm0.03$
\\
\hline
\end{tabular}
\label{tb:ffe}
\end{table*}

\begin{table*} \centering
\caption{Cumulative dust-corrected column density of Fe contained in the 
main absorber, $\log N_{\mathrm{Fe},\mathrm{m}}$ (cm$^{-2}$)}
\begin{tabular}{@{}llllll@{}}
\hline
\hline
No.\rule[-5pt]{0pt}{15pt}
& A
& B
& E0
& E1
& Mean
\\
\hline
23-24\rule{0pt}{10pt}
& $15.44\pm0.18$
& $15.00\pm0.04$
& $15.03\pm0.05$
& $15.28\pm0.14$
& $15.19\pm0.21$
\\
23-28
& $15.54\pm0.17$
& $15.15\pm0.03$
& $15.16\pm0.04$
& $15.39\pm0.18$
& $15.31\pm0.19$
\\
\hline
\end{tabular}
\label{tb:column}
\end{table*}

Since only the total column density of hydrogen atoms contained in the 
main absorber is known, Eq.~(\ref{eq:absolute}) cannot be used to 
calculate the intrinsic absolute abundances for individual components. 
Nonetheless, by cumulating the individual dust-corrected column densities 
$N_{Y,\mathrm{m}}=\sum_j N_{Y,j}/(1-f_{Y,j})$, we can calculate an 
average intrinsic metallicity $[Y/\mathrm{H}]_\mathrm{m}$. Inserting the 
calculated fractions of Fe contained in dust (Table~\ref{tb:column}) 
yields an almost solar intrinsic metallicity of 
$[\mathrm{Fe}/\mathrm{H}]_\mathrm{m}=-0.08\pm0.19$. Assuming that the 
observed \ion{H}{i} absorption is only constituted by components 23-24, 
the dust-to-metal ratio of $\rho=1.02\pm0.02$ and the intrinsic 
metallicity of $[\mathrm{Fe}/\mathrm{H}]_\mathrm{m}=-0.20\pm0.21$ give an 
average dust-to-gas ratio of $\kappa=0.64\pm0.31$.

%

\subsection{Kinematic structure}
\label{sc:kin}

With an absorption velocity interval extending for $700$~km\,s$^{-1}$ the 
kinematic distribution of associated metal line components is quite 
unique. Only the $z=1.97$ DLA system toward QSO~0013--004 shows an even 
more extended spread \citep{PetitjeanSL_2002}. The $z=2.19$ sub-DLA 
system toward HE~0001--2340 has a comparable neutral hydrogen column 
density, but a less extended absorption velocity interval of 
$400$~km\,s$^{-1}$ and much lower metallicity \citep{RichterLPB_2005}.
Rotating disks models \citep{ProchaskaJ_WolfeA_1997} and simulations of 
merging protogalactic clumps \citep{HaehneltSR_1998} do explain the 
characteristic kinematic features like asymmetric edge-leading line 
profiles, but fail to reproduce absorption intervals exceeding 
$250$~km\,s$^{-1}$. Large absorption intervals have therefore been 
associated with interacting or merging galaxies producing extended tidal 
filaments like the Antennae \citep[e.g.][]{WilsonSMC,ZhangFW_2001}. 
Another viable scenario is a line-of-sight intercepting a cluster of 
galaxies. The kinematic distribution of metal line components associated 
with the present sub-DLA system (Fig.~\ref{fg:metals}) indeed supports 
the idea that two different absorbers are involved.

\subsubsection{Peripheral components}
The peripheral components show markedly different characteristics.
For the velocity region from $-560$
to $-260$~km\,$^{-1}$ the average number density of components is about
one component per $20$~km\,s$^{-1}$. The dominating substructure is 
edge-leading, but the remaining features appear randomly distributed.
In contrast, the velocity region from $-260$ to $-60$~km\,s$^{-1}$
includes only four components, which are arranged in three isolated
groups.

\subsubsection{Main components}
The main structure is characterized by the highest frequency of peaks, 
with an average of one peak every $15$~km\,s$^{-1}$. Two substructures 
may be recognized, both edge-leading for the less refractory elements 
like Mg and Si, but rather unordered otherwise.

Particularly instructive is the comparison of extragalactic \ion{Ca}{ii} 
absorption lines with those originating in the Galactic halo 
\citep{BowenD_1991}. First ignoring components 23 and 24, the redshifted 
and the Galactic line profiles are remarkably similar 
(Fig.~\ref{fg:caii}), indicating that components 25-28 correspond to 
halo-like structures. Further developing this analogy, the narrow 
structures 23 and 24 not present in the absorption by the Galactic halo 
may be interpreted as signature of disk-like agglomeration, a picture 
which conforms with the observed depletion of elements into dust 
(Figs.~\ref{fg:gas}, \ref{fg:rho}).

\subsection{Physical conditions}
\label{sc:phys}

The physical conditions in DLA absorbers like the number density and 
kinetic temperature of hydrogen atoms and the local microwave and FUV 
radiation can be inferred from the diagnostics of fine-structure 
absorption lines \citep{BahcallJ_WolfR_1968,SilvaA_ViegasS_2002}. For the 
present absorber the analysis of excited \ion{C}{i} lines associated with 
the H$_2$ bearing components 23 and 24 provides an upper limit on the FUV 
input \citep{QuastBR_2002}. The study of H$_2$ lines predicts a radiation 
input exceeding the Galactic interstellar energy density by more than 
an order of magnitude and a number density greater than 100 hydrogen 
atoms cm$^{-2}$ \citep{ReimersBQL_2003,HirashitaH_FerraraA_2005}.  



\begin{figure*}
\centering
\includegraphics{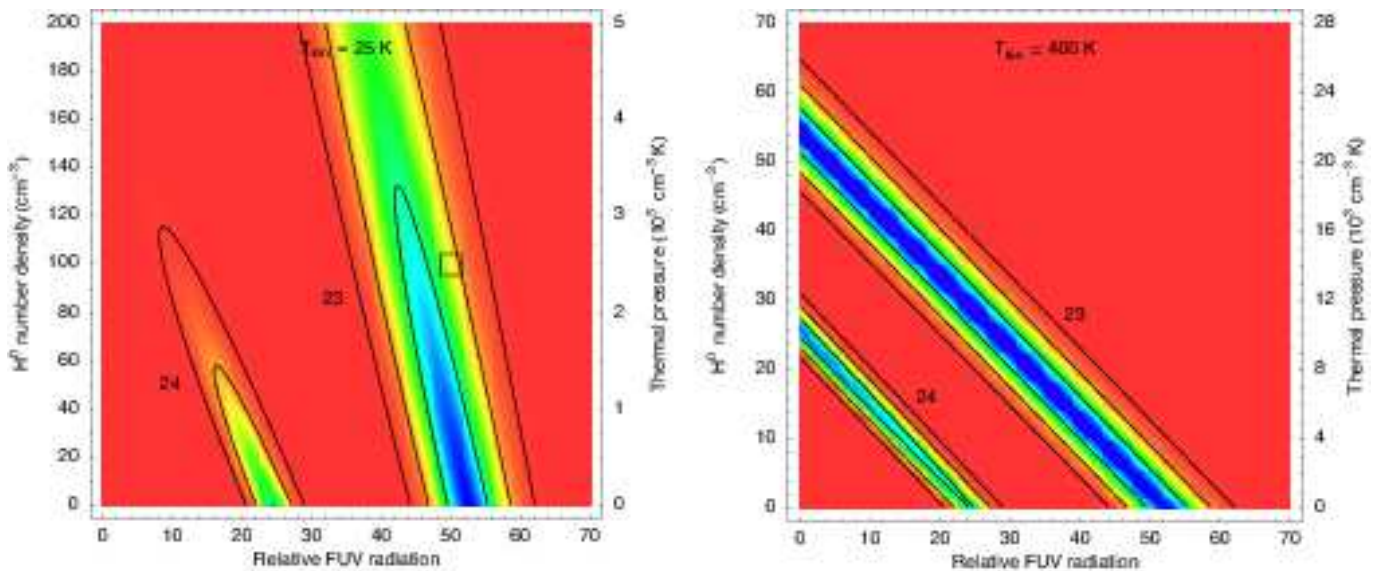}
\caption[]{Physical conditions for the H$_2$-bearing components 23 and 24 
inferred from the observed population of C\ensuremath{^0} fine-structure 
levels. Contour lines are drawn at 0.61, 0.14, and 0.01 of the maximum 
probability density and would correspond to the boundaries of the 68.3, 
95.4, and 99.7 percent confidence regions if the distributions were 
normal. Two extreme cases are considered for the kinetic gas temperature.
\emph{Left:}
The high number density of hydrogen atoms and the intense FUV radiation
found by \citet[marked by an empty square]{HirashitaH_FerraraA_2005}
require a low kinetic temperature.
\emph{Right:}
If the population of lower and higher rotational H$_2$ levels is in 
thermal equilibrium with $T_\mathrm{kin}=400$~K, either an intense FUV 
field or a high number density is ruled out}
\label{fg:phys}
\end{figure*}

Assuming an FUV input equal to the scaled generic Galactic radiation 
field \citep{DraineB_BertoldiF_1996}, but otherwise repeating the 
statistical equilibrium calculations of \citet{QuastBR_2002}, we note 
that both an intense radiation field and a number density exceeding 100 
hydrogen atoms cm$^{-2}$ only conform with the observed population of 
fine-structure levels, if the kinetic temperature is about 
$T_\mathrm{kin}=25$~K, which is different from the population 
temperature of $80$~K inferred from the lower rotational H$_2$ levels 
\citep{ReimersBQL_2003}. On the other hand, if the population of lower 
and higher rotational H$_2$ levels is in thermal equilibrium with 
$T_\mathrm{kin}=400$~K, the number density of hydrogen atoms can only 
exceed $60$~cm$^{-2}$ if the local and Galactic interstellar radiation 
are  comparable (Fig.~\ref{fg:phys}). The present spectroscopy of H$_2$ 
lines, however, is too inadequate to provide reliable results. Besides, 
the spatial distributions of carbon atoms and hydrogen molecules may not 
be identical, allowing different kinetic temperatures for both 
constituents \citep{SpitzerL_JenkinsE_1975}.
DLA systems where the population of lower and higher rotational $H_2$ levels
are not in thermal equilibrium and shielding effects are likely to play an
improtant role have been investigated by 
\citet{NoterdaemeLPLS_2007,NoterdaemePSLL_2007}.

\section{Summary and conclusions} 

Based on high-resolution spectra obtained with STIS and the VLT UVES we 
have presented a reanalysis of the chemical composition, kinematic 
structure, and physical conditions of the H$_2$-bearing sub-DLA system 
toward HE~0515--4414. The sub-damped system is unusual in several aspects:
\begin{enumerate}
  \item The velocity interval of associated metal lines extends for 
  $700$~km\,s$^{-1}$. The velocity distribution of metal line components
  is bimodal, indicating the presence of two interacting absorbers.
  \item Most of the associated metal line components are formed within
  \ion{H}{ii} regions, 
  only one third of the components associated with the predominantly
  neutral main absorber.
  \item For the main components 23-28 the observed abundance ratios of 
  refractory elements Si, Cr, Mn Fe, Ni to Zn show a distinct gradient 
  along the sight line. The differential depletion of refractory elements 
  ranges from Galactic warm disk to halo-like and essentially undepleted 
  patterns. The variation in the dust-to-metal ratio indicates the 
  multiphase structure of the absorbing medium. The dust-corrected metal 
  abundances show the nucleosynthetic odd-even effect and might imply an 
  anomalous depletion of Si relative to Fe, but otherwise do correspond 
  to solar abundance ratios. The intrinsic average metallicity is almost 
  solar, $[\mathrm{Fe}/\mathrm{H}]_\mathrm{m}=-0.08\pm0.19$, whereas the 
  uncorrected average is
  $[\mathrm{Zn}/\mathrm{H}]_\mathrm{g}=-0.38\pm0.04$. For the 
  H$_2$-bearing components 23 and 24 the dust-to-metal and dust-to-gas 
  ratios (relative to Galactic warm disk ratios) are $\rho=1.02\pm0.02$ 
  and $\kappa=0.64\pm0.31$, respectively. The ion abundances in the 
  periphery conform with solar element composition.
  \item The diagnostics of fine-structure lines is not conclusive. 
  Adequate recordings of the H$_2$ lines are needed to provide reliable 
  results.
\end{enumerate}

The presence of \ion{H}{ii} regions might have consequences for the DLA 
abundance diagnostics in general. If any metal line components are 
connected with \ion{H}{ii} rather than \ion{H}{i} regions, the usual 
averaging of element abundances is incorrect. In particular, ionization 
effects can pretend an enrichment of $\alpha$ elements. We have obtained 
a diagnostic diagram (Fig.~\ref{fg:twocolor}) which allows to detect 
\ion{H}{ii} region-like ionization conditions from empirical 
\ion{Al}{ii}, \ion{Al}{iii}, and \ion{Fe}{ii} column densities. If both 
$N_\ion{Al}{iii}/N_\ion{Al}{ii}>0.25$ and 
$N_\mathrm{Al}/N_\ion{Fe}{ii}>0.40$, the absorbing material is largely 
ionized. In this context, it is interesting to note that high column 
densities can be attained by the interception of relatively compact 
regions. For the present case \citet{ReimersBQL_2003} have already 
pointed out that the absorption path length contributed by the 
H$_2$-bearing components 23 and 24 is less than 1~lyr when the number 
density of hydrogen atoms is about $100$~cm$^{-3}$.

Our analysis shows that sub-DLA systems can exhibit solar metallicities. 
If the highest-metallicity sub-DLA absorbers prove to be regular DLA 
absorbers having consumed large amounts of neutral hydrogen due to 
massive star formation, their detection is important. Modern surveys of 
DLA systems setting the cut-off below the traditional column density 
limit may provide interesting insights.

\begin{acknowledgements}
It is a pleasure to thank Francesco Haardt for providing us with machine 
readable lookup tables of the cosmic UV background. This research has 
been supported by the Ver\-bund\-forschung of the BMBF/DLR under Grant 
No.~50\,OR\,9911\,1 and by the DFG under Re~353/48.
\end{acknowledgements}

\bibliographystyle{aa}
\bibliography{abbrev,astron,num,rq}

\Online
\small

\setcounter{figure}{1}
\begin{figure*}[p] \centering
\includegraphics{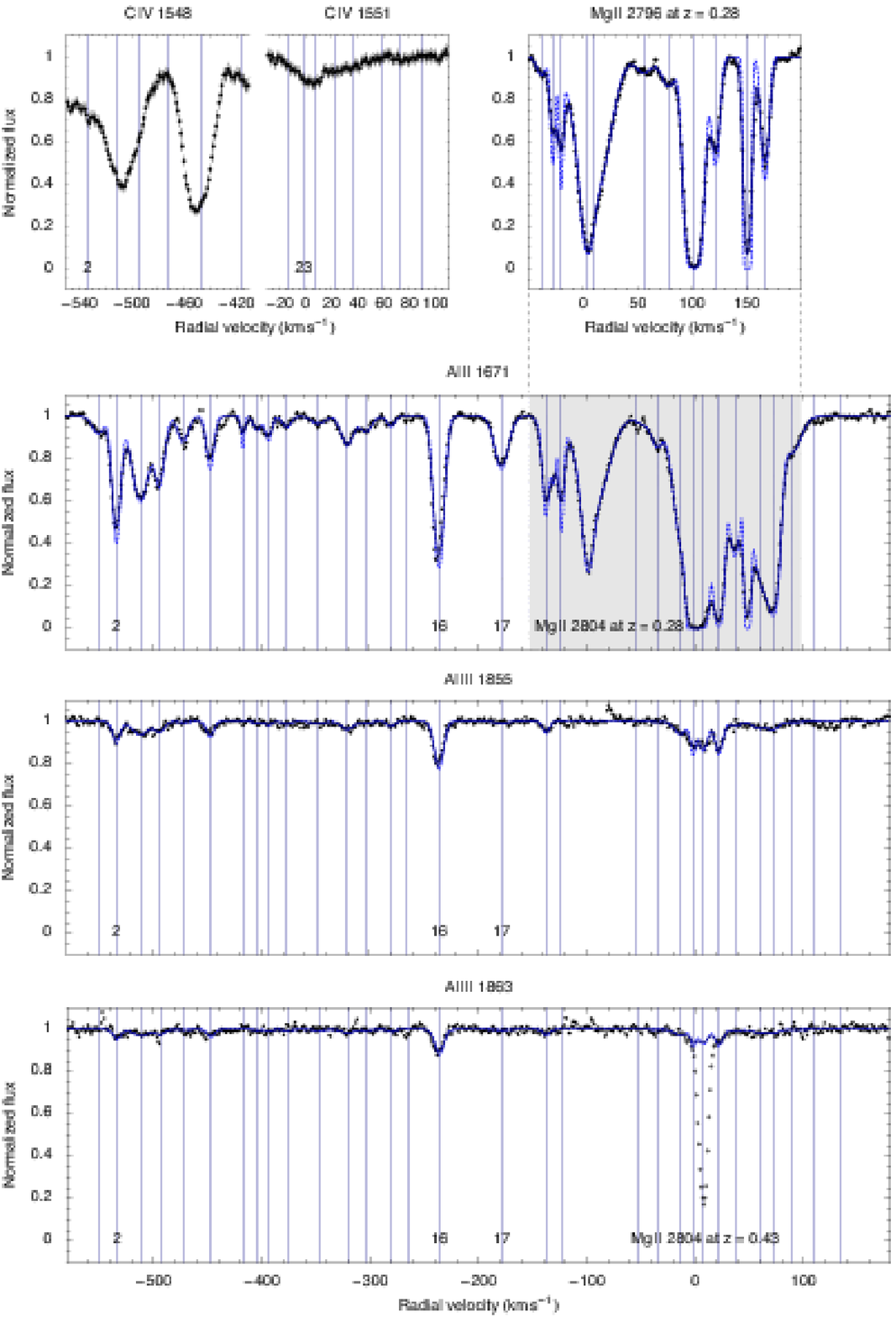}
\caption[]{continued. Note that the \ion{Al}{ii} profile is blended with 
an \ion{Mg}{ii} ensemble at redshift $z=0.28$. Blended parts of the 
\ion{C}{iv} profiles are omitted for convenience}
\label{fg:metals2}
\end{figure*}

\setcounter{figure}{1}
\begin{figure*}[p] \centering
\includegraphics{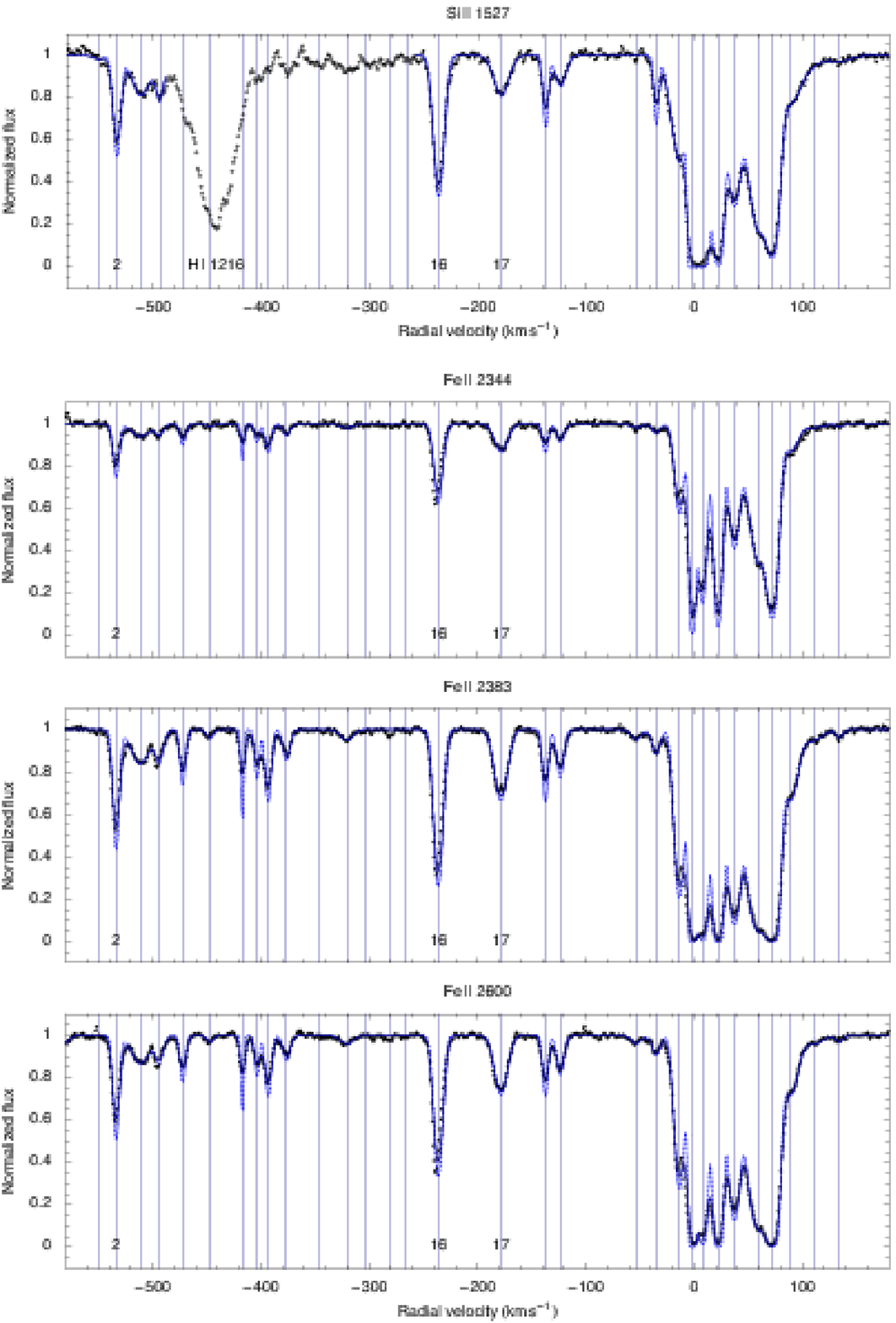}
\caption[]{continued.}
\end{figure*}

\setcounter{figure}{4}
\begin{figure*}[p] \centering
\includegraphics{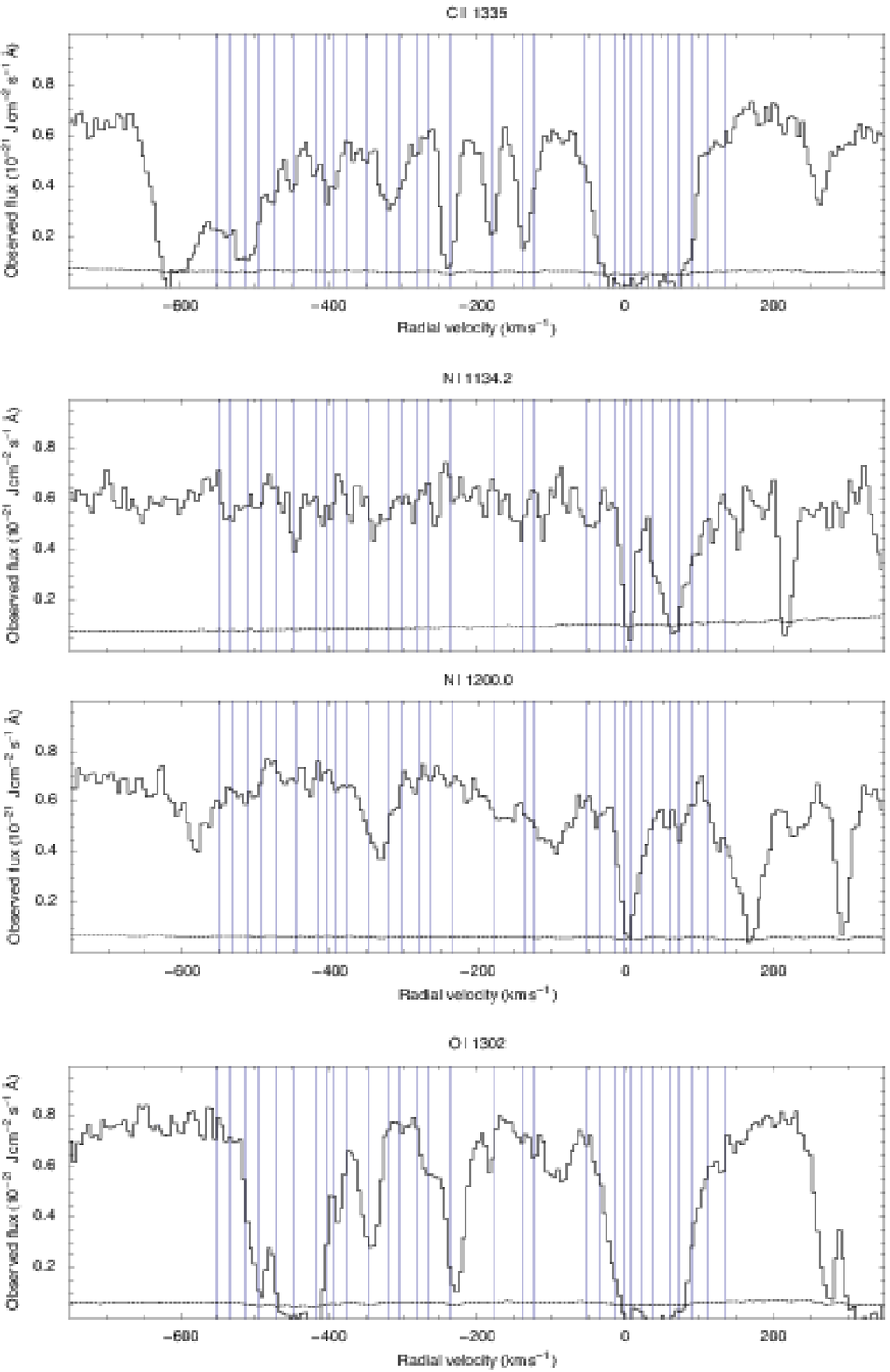}
\caption[]{Associated metal lines recorded with the STIS spectrograph. 
The solid and dashed lines mark the observed flux and its standard 
deviation, respectively. Positions of associated metal lines found with 
UVES are indicated by vertical lines. Note that many lines are blended 
within the Lyman forest}
\label{fg:stis}
\end{figure*}

\setcounter{figure}{4}
\begin{figure*}[p] \centering
\includegraphics{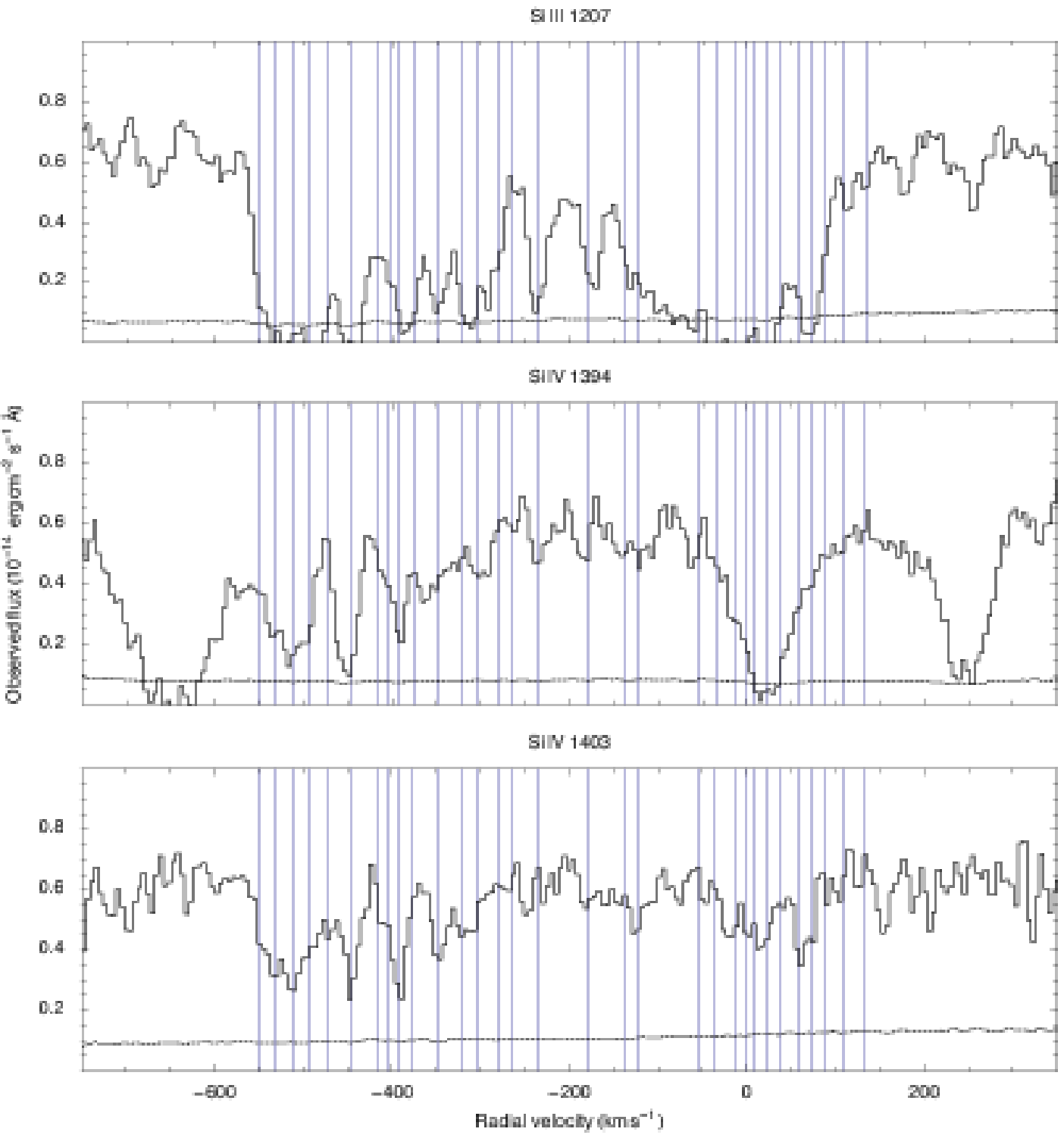}
\caption[]{continued.}
\end{figure*}

\setcounter{table}{2}
\begin{longtable}{@{}llllll@{}}
\caption[]{Optimized decomposition of the metal lines shown in 
Figs.~\ref{fg:metals} and \ref{fg:closeup}}\\
\hline
\hline
No.\rule[-5pt]{0pt}{15pt}
& \multicolumn{2}{l}{Transitions}
& $\phantom{-}\mathrm{RV}$ (km\,s$^{-1}$)
& $b$ (km\,s$^{-1}$)
& $\log N$ (cm$^{-2}$)
\\
\hline
\endfirsthead
\caption[]{continued.}\\
\hline
\hline
No.\rule[-5pt]{0pt}{15pt}
& \multicolumn{2}{l}{Transitions}
& $\phantom{-}\mathrm{RV}$ (km\,s$^{-1}$)
& $b$ (km\,s$^{-1}$)
& $\log N$ (cm$^{-2}$)
\\
\hline
\endhead
\hline
\endfoot
1\rule{0pt}{10pt}
& \ion{Mg}{ii}
& 2796, 2804
& $-550.20 \pm 0.15$
& $6.59 \pm 0.25$
& $11.79 \pm 0.02$
\\
1
& \ion{Al}{ii}
& 1671
& $-550.20$
& $7.26 \pm 1.39$
& $11.17 \pm 0.08$
\\
1
& \ion{Si}{ii}
& 1527, 1808
& $-550.20 \pm 0.15$
& $6.59 \pm 0.25$
& $11.61 \pm 0.30$
\\
\\
2
& \ion{Mg}{i}
& 2026, 2853
& $-533.51 \pm 0.03$
& $5.21 \pm 0.34$
& $11.08 \pm 0.02$
\\
2
& \ion{Mg}{ii}
& 2796, 2804
& $-533.51 \pm 0.03$
& $5.12 \pm 0.05$
& $12.90 \pm 0.01$
\\
2
& \ion{Al}{ii}
& 1671
& $-533.5$
& $4.79 \pm 0.25$
& $12.00 \pm 0.01$
\\
2
& \ion{Al}{iii}
& 1855, 1863
& $-533.5$
& $4.79 \pm 0.25$
& $11.51 \pm 0.07$
\\
2
& \ion{Si}{ii}
& 1527, 1808
& $-533.51 \pm 0.03$
& $4.45 \pm 0.37$
& $12.97 \pm 0.03$
\\
2
& \ion{Fe}{ii}
& 1608, 2344, 2374, 2383, 2587, 2600
& $-533.51 \pm 0.03$
& $3.73 \pm 0.10$
& $12.43 \pm 0.01$
\\
\\
3
& \ion{Mg}{i}
& 2026, 2853
& $-511.33 \pm 0.07$
& $\phantom{0}9.24 \pm 0.75$
& $10.94 \pm 0.04$
\\
3
& \ion{Mg}{ii}
& 2796, 2804
& $-511.33 \pm 0.07$
& $\phantom{0}9.96 \pm 0.14$
& $12.83 \pm 0.01$
\\
3
& \ion{Al}{ii}
& 1671
& $-511.33$
& $10.03 \pm 0.45$
& $12.08 \pm 0.01$
\\
3
& \ion{Al}{iii}
& 1855, 1863
& $-511.33$
& $10.03 \pm 0.45$
& $11.63 \pm 0.07$
\\
3
& \ion{Si}{ii}
& 1527, 1808
& $-511.33 \pm 0.07$
& $11.20 \pm 1.53$
& $12.89 \pm 0.05$
\\
3
& \ion{Fe}{ii}
& 1608, 2344, 2374, 2383, 2587, 2600
& $-511.33 \pm 0.07$
& $11.06 \pm 0.33$
& $12.24 \pm 0.01$
\\
\\
4
& \ion{Mg}{i}
& 2026, 2853
& $-493.59 \pm 0.08$
& $6.86 \pm 0.85$
& $10.76 \pm 0.05$
\\
4
& \ion{Mg}{ii}
& 2796, 2804
& $-493.59 \pm 0.08$
& $6.80 \pm 0.12$
& $12.53 \pm 0.01$
\\
4
& \ion{Al}{ii}
& 1671
& $-493.59$
& $6.00 \pm 0.46$
& $11.73 \pm 0.03$
\\
4
& \ion{Al}{iii}
& 1855, 1863
& $-493.59$
& $6.00 \pm 0.46$
& $11.31 \pm 0.10$
\\
4
& \ion{Si}{ii}
& 1527, 1808
& $-493.59 \pm 0.08$
& $5.43 \pm 1.36$
& $12.61 \pm 0.08$
\\
4
& \ion{Fe}{ii}
& 1608, 2344, 2374, 2383, 2587, 2600
& $-493.59 \pm 0.08$
& $5.06 \pm 0.29$
& $11.86 \pm 0.02$
\\
\\
5
& \ion{Mg}{i}
& 2026, 2853
& $-472.09 \pm0.06$
& $6.88 \pm 0.81$
& $10.62 \pm 0.05$
\\
5
& \ion{Mg}{ii}
& 2796, 2804
& $-472.09 \pm0.06$
& $5.24 \pm 0.13$
& $12.19 \pm 0.01$
\\
5
& \ion{Al}{ii}
& 1671
& $-472.09$
& $6.36 \pm 0.91$
& $11.29 \pm 0.06$
\\
5
& \ion{Al}{iii}
& 1855, 1863
& $-472.09$
& $6.36 \pm 0.91$
& $10.36 \pm 0.69$
\\
5
& \ion{Fe}{ii}
& 1608, 2344, 2374, 2383, 2587, 2600
& $-472.09 \pm0.06$
& $3.12 \pm 0.19$
& $11.92 \pm 0.02$
\\
\\
6
& \ion{Mg}{i}
& 2026, 2853
& $-447.00 \pm 0.08$
& $6.54 \pm 0.87$
& $10.50 \pm 0.06$
\\
6
& \ion{Mg}{ii}
& 2796, 2804
& $-447.00 \pm 0.08$
& $7.30 \pm 0.13$
& $12.29 \pm 0.01$
\\
6
& \ion{Al}{ii}
& 1671
& $-447.00$
& $4.46 \pm 0.44$
& $11.47 \pm 0.03$
\\
6
& \ion{Al}{iii}
& 1855, 1863
& $-447.00$
& $4.46 \pm 0.44$
& $11.35 \pm 0.09$
\\
6
& \ion{Fe}{ii}
& 1608, 2344, 2374, 2383, 2587, 2600
& $-447.00 \pm 0.08$
& $3.81 \pm 1.21$
& $11.19 \pm 0.07$
\\
\\
7
& \ion{Mg}{i}
& 2026, 2853
& $-416.85 \pm 0.06$
& $5.00 \pm 1.18$
& $10.43 \pm 0.08$
\\
7
& \ion{Mg}{ii}
& 2796, 2804
& $-416.85 \pm 0.06$
& $3.29 \pm 0.17$
& $11.87 \pm 0.01$
\\
7
& \ion{Al}{ii}
& 1671
& $-416.85$
& $1.88 \pm 1.45$
& $10.85 \pm 0.10$
\\
7
& \ion{Fe}{ii}
& 1608, 2344, 2374, 2383, 2587, 2600
& $-416.85 \pm 0.06$
& $1.72 \pm 0.20$
& $11.91 \pm 0.02$
\\
\\
8
& \ion{Mg}{ii}
& 2796, 2804
& $-403.63 \pm 0.09$
& $2.63 \pm 0.22$
& $11.78 \pm 0.02$
\\
8
& \ion{Al}{ii}
& 1671
& $-403.63$
& $4.68 \pm 1.96$
& $10.87 \pm 0.12$
\\
8
& \ion{Al}{iii}
& 1855, 1863
& $-403.63$
& $4.68 \pm 1.96$
& $10.48 \pm 0.27$
\\
8
& \ion{Fe}{ii}
& 1608, 2344, 2374, 2383, 2587, 2600
& $-403.63 \pm 0.09$
& $2.82 \pm 0.34$
& $11.81 \pm 0.02$
\\
\\
9
& \ion{Mg}{i}
& 2026, 2853
& $-393.26 \pm 0.08$
& $6.02 \pm 0.93$
& $10.58 \pm 0.07$
\\
9
& \ion{Mg}{ii}
& 2796, 2804
& $-393.26 \pm 0.08$
& $5.56 \pm 0.21$
& $12.15 \pm 0.01$
\\
9
& \ion{Al}{ii}
& 1671
& $-393.26$
& $3.81 \pm 1.33$
& $11.04 \pm 0.08$
\\
9
& \ion{Al}{iii}
& 1855, 1863
& $-393.26$
& $3.81 \pm 1.33$
& $10.66 \pm 0.30$
\\
9
& \ion{Fe}{ii}
& 1608, 2344, 2374, 2383, 2587, 2600
& $-393.26 \pm 0.08$
& $3.63 \pm 0.19$
& $12.13 \pm 0.01$
\\
\\
10
& \ion{Mg}{ii}
& 2796, 2804
& $-376.31 \pm 0.13$
& $6.02 \pm 0.26$
& $11.80 \pm 0.02$
\\
10
& \ion{Al}{ii}
& 1671
& $-376.31$
& $5.41 \pm 1.72$
& $10.83 \pm 0.12$
\\
10
& \ion{Al}{iii}
& 1855, 1863
& $-376.31$
& $5.41 \pm 1.72$
& $10.70 \pm 0.30$
\\
10
& \ion{Fe}{ii}
& 1608, 2344, 2374, 2383, 2587, 2600
& $-376.31 \pm 0.13$
& $4.37 \pm 0.24$
& $11.77 \pm 0.02$
\\
\newpage
11\rule{0pt}{10pt}
& \ion{Mg}{ii}
& 2796, 2804
& $-347.31 \pm 0.23$
& $8.56 \pm 0.38$
& $11.64 \pm 0.02$
\\
11
& \ion{Al}{ii}
& 1671
& $-347.31$
& $8.36 \pm 4.21$
& $10.85 \pm 0.15$
\\
11
& \ion{Al}{iii}
& 1855, 1863
& $-347.31$
& $8.36 \pm 4.21$
& $10.80 \pm 0.37$
\\
\\
12
& \ion{Mg}{ii}
& 2796, 2804
& $-320.68 \pm 0.11$
& $7.41 \pm 0.22$
& $12.07 \pm 0.01$
\\
12
& \ion{Al}{ii}
& 1671
& $-320.68$
& $7.41 \pm 0.95$
& $11.41 \pm 0.05$
\\
12
& \ion{Al}{iii}
& 1855, 1863
& $-320.68$
& $7.41 \pm 0.95$
& $11.26 \pm 0.12$
\\
12
& \ion{Fe}{ii}
& 1608, 2344, 2374, 2383, 2587, 2600
& $-320.68 \pm 0.11$
& $7.80 \pm 0.70$
& $11.55 \pm 0.04$
\\
\\
13
& \ion{Mg}{ii}
& 2796, 2804
& $-303.41 \pm 0.32$
& $6.59 \pm 0.43$
& $11.44 \pm 0.03$
\\
13
& \ion{Al}{ii}
& 1671
& $-303.41$
& $7.34 \pm 1.39$
& $11.11 \pm 0.09$
\\
13
& \ion{Al}{iii}
& 1855, 1863
& $-303.41 \pm 0.32$
& $7.34 \pm 1.39$
& $10.64 \pm 0.30$
\\
\\
14
& \ion{Mg}{ii}
& 2796, 2804
& $-280.56 \pm 0.15$
& $3.19 \pm 0.31$
& $11.46 \pm 0.02$
\\
14
& \ion{Al}{ii}
& 1671
& $-280.56$
& $4.45 \pm 1.37$
& $10.73 \pm 0.11$
\\
14
& \ion{Al}{iii}
& 1855, 1863
& $-280.56$
& $4.45 \pm 1.37$
& $10.87 \pm 0.20$
\\
\\
15
& \ion{Mg}{ii}
& 2796, 2804
& $-265.46 \pm 0.38$
& $4.31 \pm 0.71$
& $11.11 \pm 0.05$
\\
\\
16
& \ion{Mg}{i}
& 2026, 2853
& $-236.14 \pm 0.02$
& $6.04 \pm 0.26$
& $11.20 \pm 0.02$
\\
16
& \ion{Mg}{ii}
& 2796, 2804
& $-236.14 \pm 0.02$
& $6.25 \pm 0.04$
& $13.23 \pm 0.01$
\\
16
& \ion{Al}{ii}
& 1671
& $-236.14$
& $5.72 \pm 0.15$
& $12.22 \pm 0.01$
\\
16
& \ion{Al}{iii}
& 1855, 1863
& $-236.14$
& $5.72 \pm 0.15$
& $11.97 \pm 0.03$
\\
16
& \ion{Si}{ii}
& 1527, 1808
& $-236.14 \pm 0.02$
& $5.95 \pm 0.17$
& $13.33 \pm 0.01$
\\
16
& \ion{Fe}{ii}
& 1608, 2344, 2374, 2383, 2587, 2600
& $-236.14 \pm 0.02$
& $5.09 \pm 0.06$
& $12.78 \pm 0.01$
\\
\\
17
& \ion{Mg}{i}
& 2026, 2853
& $-178.49 \pm 0.05$
& $6.60 \pm 0.57$
& $10.74 \pm 0.04$
\\
17
& \ion{Mg}{ii}
& 2796, 2804
& $-178.49 \pm 0.05$
& $8.44 \pm 0.09$
& $12.51 \pm 0.01$
\\
17
& \ion{Al}{ii}
& 1671
& $-178.49$
& $8.28 \pm 0.47$
& $11.73 \pm 0.02$
\\
17
& \ion{Al}{iii}
& 1855, 1863
& $-178.49$
& $8.28 \pm 0.47$
& $10.66 \pm 0.30$
\\
17
& \ion{Si}{ii}
& 1527, 1808
& $-178.49 \pm 0.05$
& $9.04 \pm 0.47$
& $12.80 \pm 0.03$
\\
17
& \ion{Fe}{ii}
& 1608, 2344, 2374, 2383, 2587, 2600
& $-178.49 \pm 0.05$
& $7.77 \pm 0.13$
& $12.44 \pm 0.01$
\\
\\
18
& \ion{Mg}{i}
& 2026, 2853
& $-137.31 \pm 0.04$
& $2.74 \pm 0.86$
& $10.44 \pm 0.06$
\\
18
& \ion{Mg}{ii}
& 2796, 2804
& $-137.31 \pm 0.04$
& $3.89 \pm 0.08$
& $12.44 \pm 0.01$
\\
18
& \ion{Al}{ii}
& 1671
& $-137.31$
& $3.96 \pm 0.36$
& $11.74 \pm 0.02$
\\
18
& \ion{Al}{iii}
& 1855, 1863
& $-137.31$
& $3.96 \pm 0.36$
& $11.18 \pm 0.11$
\\
18
& \ion{Si}{ii}
& 1527, 1808
& $-137.31 \pm 0.04$
& $3.07 \pm 0.35$
& $12.62 \pm 0.03$
\\
18
& \ion{Fe}{ii}
& 1608, 2344, 2374, 2383, 2587, 2600
& $-137.31 \pm 0.04$
& $3.12 \pm 0.16$
& $12.06 \pm 0.01$
\\
\\
19
& \ion{Mg}{i}
& 2026, 2853
& $-123.63 \pm 0.07$
& $5.39 \pm 3.06$
& $10.20 \pm 0.11$
\\
19
& \ion{Mg}{ii}
& 2796, 2804
& $-123.63 \pm 0.07$
& $5.01 \pm 0.13$
& $12.18 \pm 0.01$
\\
19
& \ion{Al}{ii}
& 1671
& $-123.63$
& $3.48 \pm 0.75$
& $11.41 \pm 0.05$
\\
19
& \ion{Si}{ii}
& 1527, 1808
& $-123.63 \pm 0.07$
& $6.55 \pm 0.58$
& $12.54 \pm 0.05$
\\
19
& \ion{Fe}{ii}
& 1608, 2344, 2374, 2383, 2587, 2600
& $-123.63 \pm 0.07$
& $4.49 \pm 0.22$
& $12.01 \pm 0.02$
\\
\\
20
& \ion{Fe}{ii}
& 1608, 2344, 2374, 2383, 2587, 2600
& $-53.77 \pm 0.52$
& $6.38 \pm 0.84$
& $11.45 \pm 0.05$
\\
\\
21
& \ion{Mg}{i}
& 2026, 2853
& $-35.06 \pm 0.16$
& $6.32 \pm 1.56$
& $10.53 \pm 0.07$
\\
21
& \ion{Al}{ii}
& 1671
& $-35.06$
& $6.19 \pm 1.02$
& $11.29 \pm 0.06$
\\
21
& \ion{Al}{iii}
& 1855, 1863
& $-35.06$
& $6.19 \pm 1.02$
& $10.11 \pm 1.37$
\\
21
& \ion{Si}{ii}
& 1527, 1808
& $-35.06 \pm 0.16$
& $2.76 \pm 0.38$
& $12.56 \pm 0.04$
\\
21
& \ion{Fe}{ii}
& 1608, 2344, 2374, 2383, 2587, 2600
& $-35.06 \pm 0.16$
& $5.08 \pm 0.45$
& $11.75 \pm 0.03$
\\
\\
22
& \ion{Mg}{i}
& 2026, 2853
& $-13.53 \pm 0.07$
& $8.52 \pm 0.51$
& $11.23 \pm 0.02$
\\
22
& \ion{Al}{ii}
& 1671
& $-13.53$
& $7.98 \pm 0.37$
& $12.16 \pm 0.02$
\\
22
& \ion{Al}{iii}
& 1855, 1863
& $-13.53$
& $7.98 \pm 0.37$
& $11.50 \pm 0.08$
\\
22
& \ion{Si}{ii}
& 1527, 1808
& $-13.53 \pm 0.07$
& $9.97 \pm 0.24$
& $13.36 \pm 0.01$
\\
22
& \ion{Ca}{ii}
& 3935
& $-13.53 \pm 0.07$
& $7.89 \pm 0.97$
& $11.16 \pm 0.04$
\\
22\rule{0pt}{10pt}
& \ion{Fe}{ii}
& 1608, 2344, 2374, 2383, 2587, 2600
& $-13.53 \pm 0.07$
& $5.79 \pm 0.10$
& $12.90 \pm 0.01$
\\
\\
23
& \ion{Mg}{i}
& 2026, 2853
& $-1.61 \pm 0.03$
& $2.16 \pm 0.05$
& $12.41 \pm 0.03$
\\
23
& \ion{Al}{ii}
& 1671
& $-1.61$
& $3.05$
& $13.86 \pm 0.17$
\\
23
& \ion{Al}{iii}
& 1855, 1863
& $-1.61$
& $3.05$
& $11.50 \pm 0.06$
\\
23
& \ion{Si}{i}
& 2515
& $-1.61 \pm 0.03$
& $2.17 \pm 2.10$
& $11.34 \pm 0.08$
\\
23
& \ion{Si}{ii}
& 1527, 1808
& $-1.61 \pm 0.03$
& $3.05 \pm 0.09$
& $14.28 \pm 0.03$
\\
23
& \ion{S}{i}
& 1807
& $-1.61 \pm 0.03$
& $1.56 \pm 1.12$
& $12.18 \pm 0.05$
\\
23
& \ion{Ca}{ii}
& 3935
& $-1.61 \pm 0.03$
& $2.15 \pm 0.10$
& $12.12 \pm 0.02$
\\
23
& \ion{Cr}{ii}
& 2056, 2062
& $-1.61 \pm 0.03$
& $2.97 \pm 0.05$
& $11.84 \pm 0.06$
\\
23
& \ion{Mn}{ii}
& 2577, 2594, 2606
& $-1.61 \pm 0.03$
& $2.97 \pm 0.05$
& $11.51 \pm 0.03$
\\
23
& \ion{Fe}{i}
& 2484, 2524
& $-1.61 \pm 0.03$
& $0.48 \pm 0.18$
& $11.30 \pm 0.06$
\\
23
& \ion{Fe}{ii}
& 1608, 2344, 2374, 2383, 2587, 2600
& $-1.61 \pm 0.03$
& $2.97 \pm 0.05$
& $13.51 \pm 0.01$
\\
23
& \ion{Ni}{ii}
& 1710, 1742, 1752
& $-1.61 \pm 0.03$
& $2.97 \pm 0.05$
& $12.48 \pm 0.04$
\\
23
& \ion{Zn}{ii}
& 2026, 2063
& $-1.61 \pm 0.04$
& $2.97 \pm 0.05$
& $11.79 \pm 0.02$
\\
\\
24
& \ion{Mg}{i}
& 2026, 2853
& $\phantom{-}7.63 \pm 0.06$
& $4.09 \pm 0.12$
& $11.90 \pm 0.01$
\\
24
& \ion{Al}{ii}
& 1671
& $\phantom{-}7.63$
& $5.36$
& $12.82 \pm 0.04$
\\
24
& \ion{Al}{iii}
& 1855, 1863
& $\phantom{-}7.63$
& $5.36$
& $11.72 \pm 0.05$
\\
24
& \ion{Si}{ii}
& 1527, 1808
& $\phantom{-}7.63 \pm 0.06$
& $5.36 \pm 0.25$
& $14.16 \pm 0.04$
\\
24
& \ion{Ca}{ii}
& 3935
& $\phantom{-}7.63 \pm 0.06$
& $3.17 \pm 0.17$
& $11.78 \pm 0.01$
\\
24
& \ion{Cr}{ii}
& 2056, 2062
& $\phantom{-}7.63 \pm 0.06$
& $4.92 \pm 0.13$
& $11.61 \pm 0.14$
\\
24
& \ion{Mn}{ii}
& 2577, 2594, 2606
& $\phantom{-}7.63 \pm 0.06$
& $4.92 \pm 0.13$
& $11.36 \pm 0.03$
\\
24
& \ion{Fe}{ii}
& 1608, 2344, 2374, 2383, 2587, 2600
& $\phantom{-}7.63 \pm 0.06$
& $4.92 \pm 0.13$
& $13.36 \pm 0.01$
\\
24
& \ion{Ni}{ii}
& 1710, 1742, 1752
& $\phantom{-}7.63 \pm 0.06$
& $4.92 \pm 0.13$
& $12.33 \pm 0.06$
\\
24
& \ion{Zn}{ii}
& 2026, 2063
& $\phantom{-}7.63 \pm 0.06$
& $4.92 \pm 0.13$
& $11.59 \pm 0.03$
\\
\\
25
& \ion{Mg}{i}
& 2026, 2853
& $\phantom{-}22.06 \pm 0.03$
& $5.68 \pm 0.14$
& $11.78 \pm 0.01$
\\
25
& \ion{Al}{ii}
& 1671
& $\phantom{-}22.06$
& $4.85 \pm 0.20$
& $12.69 \pm 0.02$
\\
25
& \ion{Al}{iii}
& 1855, 1863
& $\phantom{-}22.06$
& $4.85 \pm 0.20$
& $11.72 \pm 0.05$
\\
25
& \ion{Si}{ii}
& 1527, 1808
& $\phantom{-}22.06 \pm 0.03$
& $4.99 \pm 0.23$
& $13.89 \pm 0.03$
\\
25
& \ion{Ca}{ii}
& 3935
& $\phantom{-}22.06 \pm 0.03$
& $4.37 \pm 0.21$
& $11.74 \pm 0.01$
\\
25
& \ion{Cr}{ii}
& 2056, 2062
& $\phantom{-}22.06 \pm 0.03$
& $4.29 \pm 0.06$
& $11.77 \pm 0.07$
\\
25
& \ion{Mn}{ii}
& 2577, 2594, 2606
& $\phantom{-}22.06 \pm 0.03$
& $4.29 \pm 0.06$
& $11.34 \pm 0.03$
\\
25
& \ion{Fe}{ii}
& 1608, 2344, 2374, 2383, 2587, 2600
& $\phantom{-}22.06 \pm 0.03$
& $4.29 \pm 0.06$
& $13.53 \pm 0.01$
\\
25
& \ion{Ni}{ii}
& 1710, 1742, 1752
& $\phantom{-}22.06 \pm 0.03$
& $4.29 \pm 0.06$
& $12.37 \pm0.06$
\\
25
& \ion{Zn}{ii}
& 2026, 2063
& $\phantom{-}22.06 \pm 0.03$
& $4.29 \pm 0.06$
& $11.03 \pm 0.08$
\\
\\
26
& \ion{Mg}{i}
& 2026, 2853
& $\phantom{-}36.99 \pm 0.07$
& $5.68 \pm 0.23$
& $11.43 \pm 0.02$
\\
26
& \ion{Al}{ii}
& 1671
& $\phantom{-}37.00$
& $7.33 \pm 0.43$
& $12.26 \pm 0.02$
\\
26
& \ion{Al}{iii}
& 1855, 1863
& $\phantom{-}37.00$
& $7.33 \pm 0.43$
& $10.97 \pm 0.20$
\\
26
& \ion{Si}{ii}
& 1527, 1808
& $\phantom{-}37.00 \pm 0.07$
& $7.26 \pm 0.39$
& $13.47 \pm 0.02$
\\
26
& \ion{Ca}{ii}
& 3935
& $\phantom{-}37.00 \pm 0.07$
& $6.22 \pm 0.22$
& $11.75 \pm 0.01$
\\
26
& \ion{Cr}{ii}
& 2056, 2062
& $\phantom{-}37.00 \pm 0.07$
& $6.32 \pm 0.14$
& $11.52 \pm 0.10$
\\
26
& \ion{Mn}{ii}
& 2577, 2594, 2606
& $\phantom{-}37.00 \pm 0.07$
& $6.32 \pm 0.14$
& $10.73 \pm 0.11$
\\
26
& \ion{Fe}{ii}
& 1608, 2344, 2374, 2383, 2587, 2600
& $\phantom{-}37.00 \pm 0.07$
& $6.32 \pm 0.14$
& $13.14 \pm 0.01$
\\
26
& \ion{Ni}{ii}
& 1710, 1742, 1752
& $\phantom{-}37.00 \pm 0.07$
& $6.32 \pm 0.14$
& $12.02 \pm 0.12$
\\
26
& \ion{Zn}{ii}
& 2026, 2063
& $\phantom{-}37.00 \pm 0.07$
& $6.32 \pm 0.14$
& $10.98 \pm 0.10$
\\
\\
27
& \ion{Mg}{i}
& 2026, 2853
& $\phantom{-}59.39 \pm 0.19$
& $11.53 \pm 0.40$
& $11.72 \pm 0.02$
\\
27
& \ion{Al}{ii}
& 1671
& $\phantom{-}59.39$
& $\phantom{0}8.71 \pm 1.12$
& $12.43 \pm 0.04$
\\
27
& \ion{Al}{iii}
& 1855, 1863
& $\phantom{-}59.39$
& $\phantom{0}8.71 \pm 1.12$
& $11.29 \pm 0.13$
\\
27
& \ion{Si}{ii}
& 1527, 1808
& $\phantom{-}59.39 \pm 0.19$
& $10.96 \pm 0.50$
& $13.81 \pm 0.01$
\\
27
& \ion{Ca}{ii}
& 3935
& $\phantom{-}59.39 \pm 0.19$
& $\phantom{0}8.53 \pm 0.33$
& $11.87 \pm 0.02$
\\
27
& \ion{Cr}{ii}
& 2056, 2062
& $\phantom{-}59.39 \pm 0.19$
& $11.07 \pm 0.28$
& $11.29 \pm 0.26$
\\
27
& \ion{Mn}{ii}
& 2577, 2594, 2606
& $\phantom{-}59.39 \pm 0.19$
& $11.07 \pm 0.28$
& $11.18 \pm 0.05$
\\
27
& \ion{Fe}{ii}
& 1608, 2344, 2374, 2383, 2587, 2600
& $\phantom{-}59.39 \pm 0.19$
& $11.07 \pm 0.28$
& $13.47 \pm 0.01$
\\
27
& \ion{Ni}{ii}
& 1710, 1742, 1752
& $\phantom{-}59.39 \pm 0.19$
& $11.07 \pm 0.28$
& $12.27 \pm 0.08$
\\
27
& \ion{Zn}{ii}
& 2026, 2063
& $\phantom{-}59.39 \pm 0.19$
& $11.07 \pm 0.28$
& $11.41 \pm 0.07$
\\
28\rule{0pt}{10pt}
& \ion{Mg}{i}
& 2026, 2853
& $\phantom{-}72.54 \pm 0.06$
& $6.03 \pm 0.15$
& $11.77 \pm 0.01$
\\
28
& \ion{Al}{ii}
& 1671
& $\phantom{-}72.54$
& $6.14 \pm 0.15$
& $12.57 \pm 0.02$
\\
28
& \ion{Al}{iii}
& 1855, 1863
& $\phantom{-}72.54$
& $6.14 \pm 0.15$
& $11.17 \pm 0.13$
\\
28
& \ion{Si}{ii}
& 1527, 1808
& $\phantom{-}72.54 \pm 0.06$
& $6.45 \pm 0.18$
& $13.75 \pm 0.02$
\\
28
& \ion{Ca}{ii}
& 3935
& $\phantom{-}72.54 \pm 0.06$
& $5.52 \pm 0.13$
& $12.08 \pm 0.01$
\\
28
& \ion{Cr}{ii}
& 2056, 2062
& $\phantom{-}72.54 \pm 0.06$
& $5.81 \pm 0.09$
& $11.92 \pm 0.06$
\\
28
& \ion{Mn}{ii}
& 2577, 2594, 2606
& $\phantom{-}72.54 \pm 0.06$
& $5.81 \pm 0.09$
& $11.34 \pm 0.03$
\\
28
& \ion{Fe}{ii}
& 1608, 2344, 2374, 2383, 2587, 2600
& $\phantom{-}72.54 \pm 0.06$
& $5.81 \pm 0.09$
& $13.51 \pm 0.01$
\\
28
& \ion{Ni}{ii}
& 1710, 1742, 1752
& $\phantom{-}72.54 \pm 0.06$
& $5.81 \pm 0.09$
& $12.32 \pm 0.06$
\\
28
& \ion{Zn}{ii}
& 2026, 2063
& $\phantom{-}72.54 \pm 0.06$
& $5.81 \pm 0.09$
& $10.72 \pm 0.11$
\\
\\
29
& \ion{Mg}{i}
& 2026, 2853
& $\phantom{-}89.11 \pm 0.20$
& $10.85 \pm 0.75$
& $11.08 \pm 0.04$
\\
29
& \ion{Al}{ii}
& 1671
& $\phantom{-}89.11$
& $11.24 \pm 1.11$
& $11.69 \pm 0.04$
\\
29
& \ion{Al}{iii}
& 1855, 1863
& $\phantom{-}89.11$
& $11.24 \pm 1.11$
& $10.59 \pm 0.45$
\\
29
& \ion{Si}{ii}
& 1527, 1808
& $\phantom{-}89.11 \pm 0.20$
& $13.15 \pm 1.35$
& $13.04 \pm 0.06$
\\
29
& \ion{Ca}{ii}
& 3935
& $\phantom{-}89.11 \pm 0.20$
& $10.89 \pm 1.13$
& $11.23 \pm 0.05$
\\
29
& \ion{Fe}{ii}
& 1608, 2344, 2374, 2383, 2587, 2600
& $\phantom{-}89.11 \pm 0.20$
& $\phantom{0}9.03 \pm 0.49$
& $12.50 \pm 0.02$
\\
\\
30
& \ion{Si}{ii}
& 1527, 1808
& $\phantom{-}110.24 \pm 1.43$
& $11.46 \pm 5.77$
& $12.12 \pm 0.29$
\\
30
& \ion{Fe}{ii}
& 1608, 2344, 2374, 2383, 2587, 2600
& $\phantom{-}110.24 \pm 1.43$
& $10.94 \pm 1.93$
& $11.70 \pm 0.08$
\\
\\
31
& \ion{Si}{ii}
& 1527, 1808
& $\phantom{-}134.00 \pm 0.58$
& $13.30 \pm 8.35$
& $12.12 \pm 0.31$
\\
31
& \ion{Fe}{ii}
& 1608, 2344, 2374, 2383, 2587, 2600
& $\phantom{-}134.00 \pm 0.58$
& $\phantom{0}4.33 \pm 1.22$
& $11.22 \pm 0.07$
\\
\end{longtable}

\end{document}